\documentclass[11pt]{article}
\usepackage[preprint]{acl}
\usepackage{times}
\usepackage{latexsym}
\usepackage[T1]{fontenc}
\usepackage[utf8]{inputenc}
\usepackage{microtype}
\usepackage{graphicx}
\usepackage{booktabs}
\usepackage{array}
\usepackage{amsmath,amssymb}
\usepackage{multirow}
\usepackage{xspace}
\usepackage{xcolor}

\usepackage{algorithm}
\usepackage{algpseudocode}
\usepackage{tikz}
\usetikzlibrary{positioning}

\newcommand{\synapse}{\textsc{Spur}\xspace}
\newcommand{\docs}{\textsc{Docs-only}\xspace}
\newcommand{\central}{\textsc{Centralized}\xspace}
\newcommand{\local}{\textsc{Local-only}\xspace}
\newcommand{\concat}{\textsc{Concat}\xspace}
\newcommand{\flatm}{\textsc{Flat-majority}\xspace}
\newcommand{\clf}{\textsc{Query classifier}\xspace}

\newcommand{\sd}[1]{{\tiny\ensuremath{\pm}#1}}

\title{STUN-Bench: Truly Unseen Tools for Evaluating\\Shared Tool Experience}

\author{%
  Abhijit Chakraborty \\
  Arizona State University \\
  {\small \texttt{achakr40@asu.edu}}
  \And
  Ni Trieu \\
  Arizona State University \\
  \And
  Vivek Gupta \\
  Arizona State University \\
  {\small \texttt{vgupt140@asu.edu}}
}

\begin{document}
\maketitle

\begin{abstract}
A tool router retrieves a few candidate tools per request and lets a language model choose. Methods that share experience across organizations add past requests about each tool to what the router retrieves. StableToolBench calls some of its test groups tests on ``unseen'' tools, yet the ToolBench instruction pool, which retrievers and experience-sharing methods learn from, contains every test query and labels every gold tool of the two single-tool groups among them. We release STUN-Bench (StableToolBench, Truly UNseen), a split whose experience pool has no item that labels or names 172 held-out tools. Unseen means absent from shared experience and retriever training data, not from pretraining. On the split we find crowding: experience pooled from five simulated clients, added to an index of tool descriptions, outranks the descriptions of tools no client has used, which then often leave the candidate list. Under the Jina-v2 retriever, routing accuracy on held-out tools falls from 0.534 to 0.278. Retrievers trained for tool retrieval lose much less. The split also reverses a conclusion: with every test tool labeled in the pool and our default leak filter, shared experience beats descriptions alone; on STUN-Bench it falls below them. Split, filter and harness are public\footnote{https://abhijit85.github.io/STUN-Bench/}.
\end{abstract}

\section{Introduction}

\begin{figure}[t]
\centering
\includegraphics[width=\columnwidth]{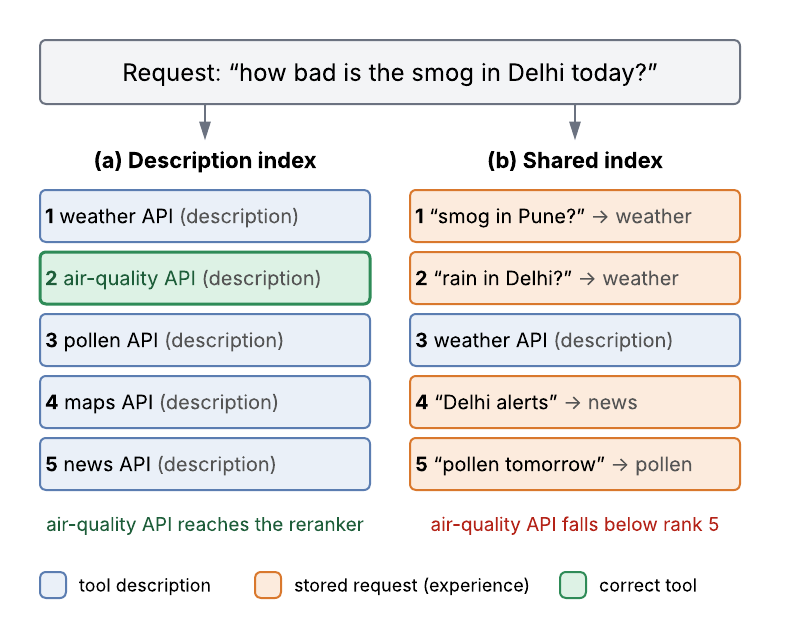}
\caption{Crowding, as an illustration (not real data). The air-quality API is new and has only its description. (a) Searching descriptions puts it among the five candidates. (b) Adding stored requests about used tools (quoted, with the tool each describes) to the same index fills the five candidates, and the reranker never sees the right tool. STUN-Bench measures how often this happens (\S\ref{sec:crowding}).}
\label{fig:example}
\end{figure}

For illustration, suppose several organizations share what their agents learned about a catalog of web APIs. A weather API has hundreds of stored requests such as ``will it rain in Pune tomorrow''. An air-quality API was added last week and has only its description. A user asks ``how bad is the smog in Delhi today''. If the stored weather requests and the tool descriptions sit in one index, the retriever may rank many weather requests above the air-quality description, and the router never sees the right tool (Fig.~\ref{fig:example}). Whether this happens, and how often, is a question about tools that no participant has used. Evaluating it needs test tools that are absent from the shared experience.

\paragraph{The standard benchmark cannot pose that question.} StableToolBench \cite{guo2024} keeps ToolBench's Tool and Category test groups, which ToolBench defines as tests on ``unseen'' tools \cite{qin2024}. ToolBench's released retriever trains on the companion ToolBench instruction pool, and derived corpora pass that pool on to other retrievers. The same pool is a ready source of simulated experience for methods that share agent memory \cite{gao2024,kim2026,chang2026,feng2026,fang2026}, prompts and exemplars \cite{chen2025fedtextgrad,wang2025fedicl}, distilled insights \cite{yao2026fot}, or concatenated ``knowledge compendiums'' for tool learning \cite{wu2024fical}. We find that the raw pool contains every StableToolBench test query and other generated queries for every gold tool of the two G1 ``unseen'' groups (\S\ref{sec:audit}). Removing duplicated test queries therefore leaves the tools labeled. After our leak filter removes exact and near-duplicate test queries, a TF-IDF classifier fit on client data still beats every LLM routing arm, on the ``unseen'' groups too. Tool-level exposure also raises a trained retriever's measured recall: under matched training, a BGE retriever that saw the held-out tools' items scores 5.6 recall@5 points higher on them (\S\ref{sec:split}).

\begin{figure*}[t]
\centering
\includegraphics[width=0.92\textwidth]{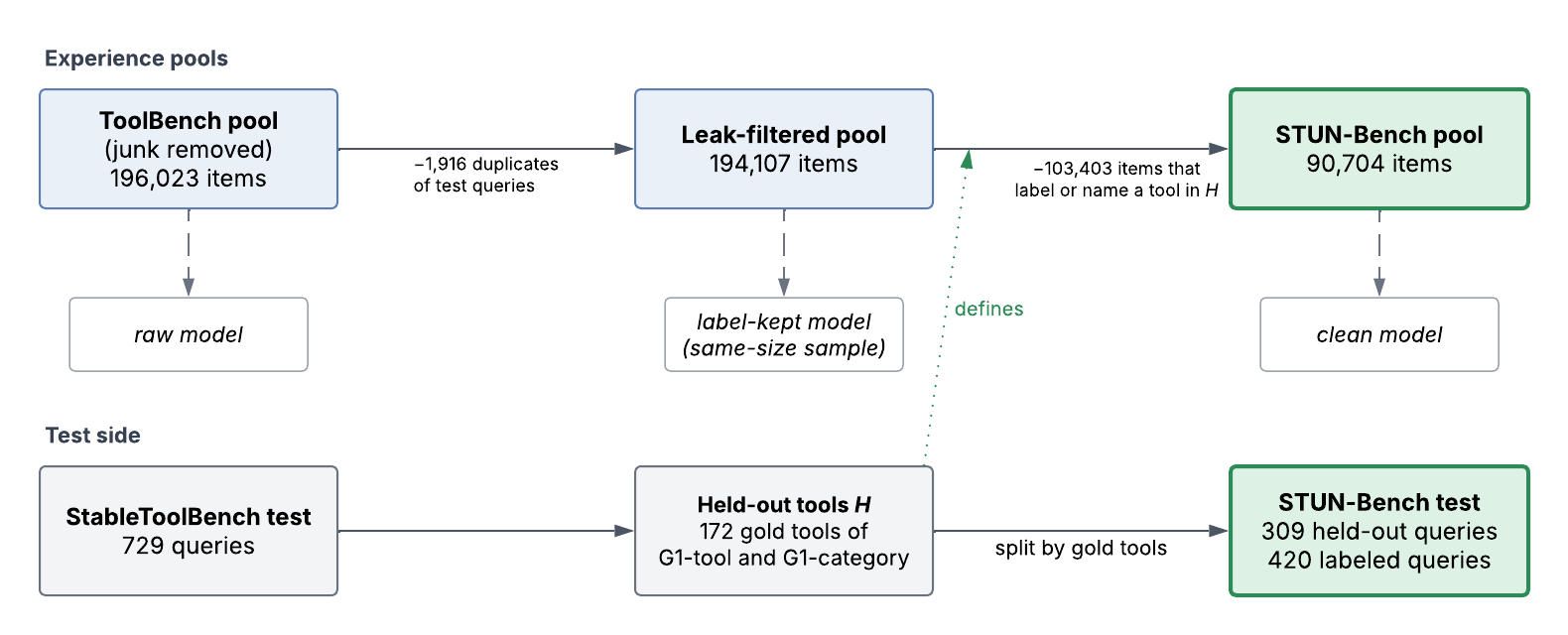}
\caption{From the ToolBench pool to STUN-Bench. Top: the experience pools; dashed arrows lead to the BGE retriever trained on each pool (\S\ref{sec:split} also uses a 90,704-item raw sample and versions filtered at cosine 0.90). Bottom: the test side. The held-out tool set $H$ defines the 103,403 removed items, 20,289 by the label rule and 83,114 by the mention rule.}
\label{fig:pipeline}
\end{figure*}

\paragraph{STUN-Bench holds out 172 tools from all shared experience.} STUN-Bench (StableToolBench, Truly UNseen) is a label-held-out split of StableToolBench. Its experience pool has no item that labels or names one of 172 held-out tools, the gold tools of the two G1 ``unseen'' groups (Fig.~\ref{fig:pipeline}). Its 309 held-out test queries have only held-out gold tools, so a supervised classifier scores zero on them by construction. ``Truly unseen'' means absent from the experience pool and from our clean retriever's training data. It does not mean absent from any base model's pretraining.

\paragraph{On the split, pooled experience crowds out unused tools.} Experience is written as user requests, so it reads more like a new request than a tool description does. In an index shared with the descriptions, tools that some client has experience with outrank tools that no client has, and the unused tools often leave the five candidates the router reranks. We call this \emph{crowding}. Under the Jina-v2 retriever and a Llama-3.1-8B reranker, adding pooled experience to the description index lowers held-out routing accuracy from 0.534 to 0.278 (\S\ref{sec:crowding}). Document expansion \cite{nogueira2019} applied to only some documents predicts the direction of this loss. We measure its size, its depth in the ranking and how it varies across retrievers.

\paragraph{The split changes what an evaluation concludes.} With every test tool labeled in the pool and our default leak filter, the same shared experience beats descriptions alone over all test queries. On STUN-Bench it falls below them. A method evaluated only on the leaky setup would report a gain that the split does not show.

\paragraph{How large the loss is depends on the retriever.} Each of four general-purpose retrievers loses held-out recall on every pool it was run on, including one with no ToolBench ancestry. Retrievers trained for tool retrieval lose 1.7 to 7.1 recall points, and for our clean trained retriever the end-to-end accuracy loss is not detected. A \emph{two-index router}, which keeps experience in its own index and fills three of five candidate slots from descriptions, recovers most of the loss under Jina-v2. With the clean retriever it trades labeled accuracy for held-out accuracy, so it is a conditional mitigation.

Our contributions are (i) an exposure audit of the ToolBench pool with a released leak filter; (ii) STUN-Bench with a datasheet (App.~\ref{app:datasheet}), and a matched measurement of how much tool-level exposure raises trained-retriever recall; and (iii) measurements of crowding across retrievers, pools, expansion and candidate depth, with the two-index router as a conditional mitigation. As a secondary result, pooled experience helps a client more than its own experience on tools only another client has used. Controls for comparing exchange formats are in App.~\ref{app:controls}. Released artifacts are listed in App.~\ref{app:artifacts}.

\begin{table}[t]
\centering\scriptsize
\setlength{\tabcolsep}{2pt}
\begin{tabular}{@{}>{\raggedright\arraybackslash}p{0.27\columnwidth}>{\raggedright\arraybackslash}p{0.66\columnwidth}r@{}}
\toprule
Term & Meaning & \S \\
\midrule
Leak-filtered pool & ToolBench pool without exact or near-duplicate test queries & \ref{sec:audit} \\
STUN-Bench & our split: no experience item labels or names 172 held-out tools & \ref{sec:split} \\
Held-out, labeled queries & test queries whose gold tools are all held out (309), and the rest (420) & \ref{sec:split} \\
Clean, label-kept, raw models & BGE-base trained on the STUN-Bench pool, a same-size sample of the leak-filtered pool, and the raw pool & \ref{sec:split} \\
Compendium & one client's experience: per tool, usage scenarios written as requests, precautions and API names & \ref{sec:protocol} \\
Description index (\docs) & tool descriptions only & \ref{sec:protocol} \\
Shared index & description index plus all clients' compendiums, merged per tool (\synapse) or pooled unmerged (\central) & \ref{sec:protocol} \\
Two-index router & three candidates from the description index, two from a separate experience index & \ref{sec:protocol} \\
\local & one client's own compendium plus descriptions, client mean & \ref{sec:protocol} \\
\clf & TF-IDF + linear SVM on the client items, no retrieval & \ref{sec:audit} \\
Crowding & experience about used tools pushing the descriptions of unused tools out of the candidates & \ref{sec:crowding} \\
\bottomrule
\end{tabular}
\caption{Terms used throughout. Accuracy is top-1 routing accuracy; recall@5 counts a gold tool among the five candidates.}
\label{tab:terms}
\end{table}

\section{Related work}

\paragraph{Tool retrieval and its benchmarks.} ToolLLM \cite{qin2024}, Gorilla \cite{patil2024} and Toolformer \cite{schick2023} select tools from documentation in a centralized setting. ToolLLM's retrieve-then-select pipeline, also used by Graph RAG-Tool Fusion \cite{lumer2025}, is the router studied here. Off-the-shelf retrievers score poorly on tool retrieval \cite{shi2025toolret,kachuee2024}, and lexical baselines are hard to beat out of domain \cite{thakur2021}. Expanding documents with predicted queries helps the documents that receive it \cite{nogueira2019}, and zero-shot tool retrievers expand every tool with generated queries \cite{chen2024reinvoke}. We test template and model-generated expansion of every tool. Documentation alone supports zero-shot use of unseen tools \cite{hsieh2023,mekala2024}, which the split tests.

\paragraph{Contamination and benchmark audits.} Verbatim and paraphrased train--test overlap inflates evaluations \cite{sainz2023,deng2023,balloccu2024,zhang2024gsm}, and string matching misses paraphrases \cite{yang2023,lee2022}. To our knowledge, audits of agentic evaluation \cite{zhu2025best,bhat2026,du2024} have not reported that the ToolBench pool contains the StableToolBench test queries or labels the ``unseen'' test tools. Tool-level exclusion has precedents. ToolRerank \cite{zheng2024} removes tools and categories from ToolBench training data, and LoSemB \cite{zhuang2026} finds that fine-tuned retrievers degrade as the share of unseen tools grows. ToolSense \cite{hathidara2026toolsense} rewrites ToolBench queries at three ambiguity levels and finds that parametric tool retrievers lose much of their recall on them while embedding and BM25 baselines do not. We hold query wording fixed and change which tools count as unseen. Cartograph \cite{agyemang2026cartograph} reports a 4-point recall@5 drop when only part of a catalog gets richer LLM-written descriptions. Crowding is a related imbalance, caused by usage experience under a tool hold-out.

\paragraph{Exchanging experience across frozen models.} Weight- and adapter-based federated learning for LLMs needs a shared base model \cite{ye2024,kuang2024,wu2024fedbiot,sun2024}, and logit distillation needs a shared dataset \cite{li2019fedmd,lin2020}. Federated RAG exchanges passages or embeddings \cite{zeng2024a,mao2025,zhao2024frag,addison2024}. For frozen, heterogeneous models the exchanged unit is text: prompts \cite{chen2025fedtextgrad,bai2026fedehr}, exemplars \cite{wang2025fedicl}, insights \cite{yao2026fot}, shared memory \cite{zhang2024survey,gao2024,kim2026,chang2026,feng2026,fang2026}, and FICAL's concatenated compendium \cite{wu2024fical}. Among these, FICAL is the one we found that indexes tool descriptions and shared experience together, which is the shared index we test. None reports a label-held-out evaluation, and a recent survey of federated agents \cite{yang2026fao} names their credible evaluation as an open problem. FICAL uses synthetic Public-API tasks, so we revise no published ranking. App.~\ref{app:paradigms} compares the paradigms.

\section{The ToolBench pool labels the ``unseen'' test tools}
\label{sec:audit}

StableToolBench \cite{guo2024} has 3,180 tools after we remove 192 junk catalog entries. Its six test groups hold 729 queries after a junk filter (765 before): G1 single-tool (instruction, tool, category), G2 intra-category multi-tool (instruction, category) and G3 intra-collection multi-tool (instruction), named I1--I3 in \citet{guo2024}. Clients draw experience from the ToolBench instruction pool: the single-tool, intra-category and intra-collection query files (196,023 items).

\paragraph{The raw pool contains the test set.} Every StableToolBench test query appears in the raw pool as an exact string match after case and whitespace normalization (729/729 after the junk filter; 765/765 before), and a TF-IDF classifier fit on the raw pool scores 1.000. ToolBench's released retriever preprocessing excludes one test-ID list, the single-tool instruction list used in its retriever example. After that exclusion, 314 of the 765 raw test queries still match pool text exactly (Tab.~\ref{tab:lineage}). We found no released step that excludes the test tools. Methods that draw from the raw pool, as we and others do, must apply both exclusions themselves. Our leak filter handles the query side for all six post-junk groups. It removes 1,916 items that duplicate a test query, exactly (745) or at cosine $\ge 0.95$ under Jina-v2 embeddings~\cite{guenther2023jina} (1,171). After it, 319/729 test queries still have a pool paraphrase at cosine $\ge 0.90$, which we report and leave in.

\paragraph{After the filter, word overlap alone beats every LLM arm.} A TF-IDF + linear SVM classifier fit on the same 25,000 labeled client items per seed scores 0.817 over the 729 queries, 23 points above the best LLM routing arm, and it leads every arm on the Tool and Category groups (Tab.~\ref{tab:groups}). The pool holds other generated queries for every gold tool of these groups (95/95 G1-tool and 77/77 G1-category). Each was written from its tool's description and shares its vocabulary. Per seed, only 1--4 of 729 test queries have a gold tool absent from the clients' items. A tighter filter at cosine $\ge 0.90$ removes 11,800 items. The refit classifier still scores 0.774\sd{.032}, about 24 points above the best LLM arm (Tab.~\ref{tab:threshold}).

\paragraph{What this setup cannot test.} Every test tool is labeled somewhere in the pool, so the setup cannot ask whether pooled experience preserves access to tools no client has used. App.~\ref{app:reuse} records what sixteen published methods that learn from ToolBench data state about exclusion: two state that they excluded the test queries, and two that they excluded the held-out tools. Retrained without the held-out tools' items, ToolBench's own retriever loses more than half its recall@5 on the Tool group, while a size-matched raw sample does not (Tab.~\ref{tab:tbir}). Our rerun stays well below the released checkpoint, so it corrects no published result.

\section{STUN-Bench: a label-held-out split}
\label{sec:split}

\paragraph{The split removes every item that labels or names a held-out tool.} The held-out tool set $H$ is the union of the gold tools of the G1-tool (95) and G1-category (77) test queries ($|H|=172$). G2-category, another ToolBench ``unseen'' group, contributes no tools to $H$. Two rules remove items from the leak-filtered pool. The \emph{label rule} removes an item if any of its labels is in $H$ (20,289 items). The \emph{mention rule} removes it if a normalized held-out tool name appears in its query, gold tool names, or attached tool, API or category metadata (83,114 items). The remaining 90,704 items form the \emph{label-held-out pool}. The 309 queries whose gold tools are all in $H$ (286 single-tool, 23 multi-tool) are \emph{held-out}. The other 420 are \emph{labeled}: 340 with no gold tool in $H$, and 80 multi-tool queries with only some of their gold tools in $H$. Each run asserts that a supervised classifier's held-out accuracy is zero. Two annotators who are not authors checked the gold labels of 50 held-out queries (App.~\ref{app:datasheet}). One gold tool could not serve its request, and 23 labels were incomplete: the gold tool covered only part of the request, or another tool served it at least as well. Dropping all 24 queries changes the Jina-v2 held-out loss of \S\ref{sec:crowding} from 25.6 to 24.8 points.

\paragraph{The label rule is the one that matters.} With the label rule alone, the shared index's held-out recall@5 is 0.321\sd{.008}; with both rules it is 0.333\sd{.006}. With the mention rule alone it is 0.775\sd{.015}, above the description index, because clients then still hold labeled queries for the tools under test. The conservative mention rule thus moves held-out recall by about one point, in favor of held-out tools. Holding out whole categories (738 tools) also produces a loss under Jina, BGE~\cite{xiao2024cpack} and BM25~\cite{robertson2009bm25}, smaller than on the split (Tab.~\ref{tab:e7}).

\paragraph{Tool-level exposure raises trained-retriever recall.} BGE-base fine-tuned on the label-held-out pool never sees a held-out tool as a target; this is the \emph{clean} model (App.~\ref{app:setup}). The \emph{label-kept} model trains on a uniform sample of the leak-filtered pool of the same size, with no test-query matches, drawn without excluding items that label or name held-out tools. Our tightest comparison fixes the filter at cosine 0.90 and gives both models 87,599 items and 1,051 updates. Held-out recall@5 from descriptions is then 0.873 for the clean model and 0.929 for the label-kept one, a 5.6-point gap with tool-clustered 95\% interval $[+3.2,+8.2]$ (Tab.~\ref{tab:bgearms}). Labels and mentions change together here, so the gap measures their joint exposure. On the independent corpus (\S\ref{sec:protocol}) the two models are within half a point (0.224 vs.\ 0.219).

\paragraph{Looser filters and raw training show the same pattern.} At the 0.95 filter with 90,704 items each, the label-kept model scores 5.2 points above the clean one (tool-clustered $[+2.8,+7.7]$). Several published tool retrievers train on the raw pool (App.~\ref{app:reuse}). Trained there, our recipe gains 8.4 points on the held-out tools (tool-clustered $[+5.1,+12.0]$; Tab.~\ref{tab:retrievers}) and none on the independent pool (0.218 vs.\ 0.219). A uniform raw sample of the same size keeps most of that gain and scores 1.4 points above the label-kept model (tool-clustered $[+0.1,+2.9]$), so in our runs pool volume accounts for little of it. Each step changes more than one thing (App.~\ref{app:setup}), so the 5.2/1.4 split compares exposure conditions and does not assign causes.

\section{Evaluation protocol}
\label{sec:protocol}

\paragraph{Router.} We retrieve candidates by cosine similarity under Jina-v2, unless stated otherwise, and walk down the ranking until five distinct tools are collected (cap 200 entries). A frozen LLM reranker reads each candidate's description, ``when to use'' text and ``do not use when'' text in a fixed description-first prompt, scores candidate indices by next-token log-probability, and routes to the argmax. Accuracy is top-1 match on normalized tool ids; on multi-tool queries a route is correct if it matches any gold tool. Recall@5 counts a hit if any gold tool is among the five candidates. Llama-3.1-8B-Instruct~\cite{grattafiori2024llama3} is the default reranker (App.~\ref{app:setup}).

\paragraph{Experience.} Five clients each receive 5,000 items from the pool, partitioned by category, and each turns its items into a compendium: canonical tool metadata, usage scenarios written as requests and tagged with a tool, precautions from observed near-misses, and API names. In the \synapse arm (Shared Pool of Usage Records) an edge aggregator clusters each tool's scenarios at cosine $\ge 0.85$ and keeps a per-field majority (App.~\ref{app:merge}); in \central the compendiums are pooled without that merge. Both arms put the result in the same index as the descriptions, which makes the shared index. The two-index router keeps it in a separate index. \local uses one client's compendium, and \clf fits the client items directly. We run three rounds with seeds 42/123/456 and report mean$\pm$SD. Intervals are paired bootstrap 95\%, clustered by query unless marked tool-clustered; pooled McNemar $p$-values are descriptive.

\paragraph{Diagnostics.} To separate the candidates from what the reranker reads, we cross candidate source with rendering on the same queries (App.~\ref{app:setup}). \emph{Gold insertion} puts every gold tool absent from a query's five candidates in place of the lowest-ranked non-gold candidate and reruns the reranker. It shows what the reranker does when retrieval never misses; it is not a reranking ceiling, which under the any-gold rule is the post-insertion recall itself.

\paragraph{Other pools.} The retrieval stage alone also runs on two more pools. The ToolBench single-tool pool has 1,606 tools, 482 held out per seed. The independent pool uses 32 tool-retrieval evaluation sets of the ToolRet benchmark \cite{shi2025toolret} with no ToolBench ancestry (5,140 tools, 350 held out per draw); it has 0 exact and 4 near-duplicate matches to the ToolBench pool (App.~\ref{app:reuse}). The full pipeline also runs on the retail domain of $\tau$-bench \cite{yao2024tau} (App.~\ref{app:tau}).

\section{Crowding on STUN-Bench}
\label{sec:crowding}

\begin{figure*}[t]
\centering
\includegraphics[width=\textwidth]{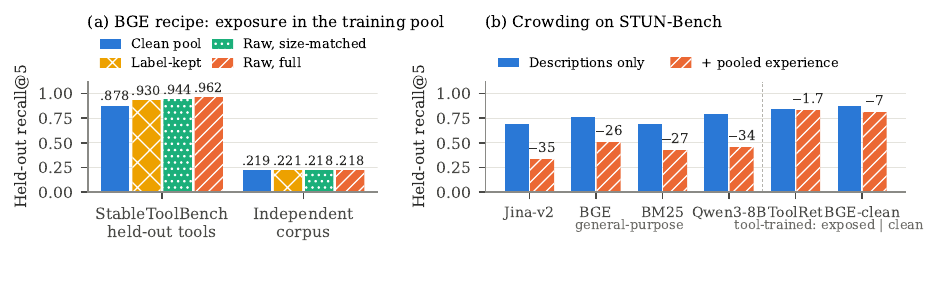}
\caption{Exposure in training raises held-out recall, and pooled experience lowers it. (a) The BGE recipe trained on four pools (\S\ref{sec:split}; clean and label-kept at the 0.95 filter, raw pools unfiltered; 0.90 results in Tab.~\ref{tab:bgearms}), on held-out tools and the independent corpus. (b) Held-out recall@5 on STUN-Bench from descriptions and the shared index; labels: loss in points. Right of the dashed line: tool-trained retrievers, exposed (ToolRet-e5) and clean. Three seeds.}
\label{fig:main}
\end{figure*}

\begin{table}[t]
\centering\scriptsize
\setlength{\tabcolsep}{1.8pt}
\begin{tabular}{@{}lcccc@{}}
\toprule
 & \multicolumn{2}{c}{held-out (309)} & \multicolumn{2}{c}{labeled (420)} \\
\cmidrule(lr){2-3}\cmidrule(lr){4-5}
Arm & acc & R@5 & acc & R@5 \\
\midrule
\multicolumn{5}{@{}l}{Jina-v2} \\
\docs & 0.534 & 0.689 & 0.524 & 0.660 \\
Shared index, \synapse & 0.278\sd{.006} & 0.333\sd{.006} & 0.528\sd{.009} & 0.659\sd{.009} \\
Two-index router & 0.520\sd{.002} & 0.614\sd{.002} & 0.553\sd{.015} & 0.724\sd{.009} \\
\central & 0.284\sd{.007} & 0.335\sd{.007} & 0.523\sd{.019} & 0.658\sd{.015} \\
\local$^{\dagger}$ & 0.410 & 0.493 & 0.398 & 0.493 \\
\clf & 0.000 & -- & 0.637\sd{.010} & -- \\
\midrule
\multicolumn{5}{@{}l}{Clean BGE retriever$^{a}$} \\
\docs & 0.687\sd{.010} & 0.878\sd{.004} & 0.706\sd{.006} & 0.930\sd{.005} \\
Shared index & 0.665\sd{.019} & 0.807\sd{.002} & 0.706\sd{.011} & 0.924\sd{.009} \\
Two-index router & 0.723\sd{.005} & 0.821\sd{.002} & 0.676\sd{.013} & 0.930\sd{.005} \\
\bottomrule
\end{tabular}
\caption{Under Jina the shared index halves held-out accuracy. Clean retriever training shrinks the loss. With the clean retriever, the two-index router trades labeled for held-out accuracy. Seeds 42/123/456. Two-index router: three description and two experience slots. Acc: top-1, correct on any gold tool. R@5: recall@5. Labeled includes 80 mixed queries (Tab.~\ref{tab:strata}); diagnostic arms in Tab.~\ref{tab:diag}. $^{\dagger}$Seeds 42/456. $^{a}$Trained on the label-held-out pool.}
\label{tab:heldout}
\end{table}

\begin{table*}[t]
\centering\small
\setlength{\tabcolsep}{2.6pt}
\begin{tabular}{@{}lcccccc@{}}
\toprule
 & \multicolumn{2}{c}{STUN-Bench} & \multicolumn{2}{c}{ToolBench G1} & \multicolumn{2}{c}{Independent} \\
\cmidrule(lr){2-3}\cmidrule(lr){4-5}\cmidrule(lr){6-7}
Retriever & docs & shared & docs & shared & docs & shared \\
\midrule
Jina-v2 & .689 & .335\sd{.007} & .763\sd{.003} & .263\sd{.005} & .169\sd{.027} & .112\sd{.020} \\
BGE-base & .764 & .502\sd{.012} & .847 & .360\sd{.009} & .181\sd{.030} & .112\sd{.031} \\
BM25 & .689 & .421\sd{.003} & .713\sd{.007} & .530\sd{.008} & .136\sd{.011} & .058\sd{.013} \\
Qwen3-Emb-8B & .790 & .454\sd{.008} & -- & -- & .117\sd{.022} & .102\sd{.021} \\
ToolBench-IR$^{\S}$ & .974 & .924\sd{.007} & -- & -- & .005\sd{.001} & .000 \\
ToolRet-e5$^{\ast}$ & .845 & .827\sd{.004} & -- & -- & -- & -- \\
BGE, clean$^{\P}$ & .878\sd{.004} & .807\sd{.002} & -- & -- & .219\sd{.039} & .216\sd{.038} \\
BGE, label-kept$^{\#}$ & .930\sd{.005} & .890\sd{.009} & -- & -- & .221\sd{.041} & .216\sd{.040} \\
BGE, raw$^{\ddagger}$ & .962\sd{.005} & .920\sd{.009} & -- & -- & .218\sd{.041} & .213\sd{.039} \\
\bottomrule
\end{tabular}
\caption{Each general-purpose retriever loses held-out recall@5 from the shared index on each pool it was run on, and tool-trained retrievers lose less. Columns: description index (docs) and shared index on the three pools of \S\ref{sec:protocol}. $^{\S}$Released ToolBench retriever. $^{\P}$Clean model; not run on ToolBench G1, whose held-out tools it was trained on. $^{\ast}$ToolRet's released E5-base tool retriever; not run on the independent pool, built from the same ToolRet benchmark. $^{\#}$Label-kept and $^{\ddagger}$raw-pool versions of the clean recipe.}
\label{tab:retrievers}
\end{table*}

\paragraph{Under Jina, pooled experience halves held-out accuracy.} From descriptions alone \docs routes 0.534 of held-out queries correctly. \synapse reaches 0.278\sd{.006} and \central, which skips the merge, 0.284\sd{.007} (Tab.~\ref{tab:heldout}; Fig.~\ref{fig:main}). The 26-point loss has tool-clustered 95\% interval $[-30.8,-20.6]$ (Tab.~\ref{tab:toolclust}). Crossing candidate source with rendering locates it: the index changes held-out accuracy by about 24 points under either rendering, and the rendering changes it by 0.7 (Tab.~\ref{tab:cross}). The shared index misses the gold tool twice as often as the description index (recall@5 0.333 vs.\ 0.689). With the gold tool inserted into the five candidates, it still scores below descriptions (0.526 vs.\ 0.644; Tab.~\ref{tab:diag}), because its other four candidates differ. On labeled queries the shared index is within half a point of descriptions (0.528 vs.\ 0.524), so the loss falls on tools without experience.

\paragraph{The loss grows with the amount of experience.} Held-out recall@5 is 0.689 with no experience, 0.493 with one client's compendium (\local) and 0.333 with all five clients' compendiums. The direction is what document expansion applied to some documents predicts \cite{nogueira2019}; the size and depth are what we measure.

\begin{table}[t]
\centering\scriptsize
\setlength{\tabcolsep}{2.5pt}
\begin{tabular}{@{}lccc@{}}
\toprule
 & Leak-filtered & \multicolumn{2}{c}{STUN-Bench} \\
\cmidrule(lr){2-2}\cmidrule(lr){3-4}
Arm & all (729) & all (729) & held-out (309) \\
\midrule
\docs & .528 & .528 & .534 \\
Shared index, \synapse & .587 & .422\sd{.003} & .278\sd{.006} \\
\clf & .817 & -- & .000 \\
\bottomrule
\end{tabular}
\caption{The split reverses the comparison of shared experience with descriptions. Routing accuracy under Jina-v2 and Llama-3.1-8B, three seeds. Leak-filtered: every test tool labeled, filter at cosine 0.95 (Tab.~\ref{tab:groups}). STUN-Bench: Tab.~\ref{tab:twoindex} and Tab.~\ref{tab:heldout}.}
\label{tab:reversal}
\end{table}

\paragraph{The split reverses a conclusion of the leaky setup.} On the leak-filtered pool, where every test tool is labeled, \synapse beats descriptions over all 729 queries (0.587 vs.\ 0.528 at the 0.95 filter). On STUN-Bench, under the same retriever and reranker, it falls below them (0.422 vs.\ 0.528; Tab.~\ref{tab:reversal}). At the tighter 0.90 filter, \synapse is already below descriptions on the leak-filtered pool (0.480 vs.\ 0.528, one seed; Tab.~\ref{tab:threshold}).

\paragraph{Each general-purpose retriever loses held-out recall.} Under Jina-v2, BGE-base, BM25 and Qwen3-Embedding-8B~\cite{zhang2025qwen3emb}, the loss appears on every pool each was run on (Tab.~\ref{tab:retrievers}), including the one with no ToolBench ancestry. It is 26--35 points on the split and 18--50 on the ToolBench single-tool pool. On the independent pool it is 1.5--8 points, and every retriever's absolute recall there is low. Qwen3-Embedding-8B reaches more held-out gold tools from descriptions than the other three (0.790) and loses 33.6 points under the shared index ($[-38.9,-28.4]$).

\paragraph{Retrievers trained for tool retrieval lose less.} The shared index costs the clean model 7.1 held-out recall points (0.878 to 0.807\sd{.002}), with labeled recall 0.6 points lower (0.930 vs.\ 0.924) and a 0.3-point loss on the independent pool. End to end, it costs 2.3 points of held-out accuracy, with a tool-clustered interval that includes zero (Tab.~\ref{tab:heldout}). ToolRet-e5, E5-base~\cite{wang2022e5} trained on ToolRet-train \cite{shi2025toolret}, loses 1.7 points (0.845 to 0.827\sd{.004}; tool-clustered $[-3.1,-0.6]$). ToolRet-train contains most of the test queries (App.~\ref{app:reuse}), so its held-out number is not a clean baseline.

\paragraph{Under tool-trained retrievers, descriptions outscore experience more often.} For a held-out query, we compare the gold description's score with that of the nearest experience entry, which describes another tool (Tab.~\ref{tab:gap}). Experience outscores the gold description for 59--78\% of held-out queries under the general-purpose retrievers and 22--36\% under the tool-trained ones, and the six retrievers fall in the same order by this share as by recall loss. Share and loss are linked by construction, so this locates the displacement and is not a separate causal test. One possible reading, which we did not test, is that training on query--description pairs lifts descriptions above query-shaped text.

\paragraph{The displacement reaches past the five candidate slots.} At a depth of 200 distinct tools the shared index reaches a held-out gold tool for 0.558\sd{.008} of queries, against 0.961 for descriptions. With gold insertion at that depth the reranker ends 25 points behind (0.400 vs.\ 0.650; Tab.~\ref{tab:bounded}).

\paragraph{Generated expansion of every tool shrinks the loss; template expansion does not.} Five template queries per tool, written from its own description, are roughly neutral alone, and the loss stays once experience is added. Model-generated expansion, as zero-shot tool retrievers use \cite{chen2024reinvoke}, lowers held-out recall on its own; pooled experience then costs 14.0 points under Jina, less than one under BGE and 17.3 under BM25, against 34.9, 17.5 and 34.1 under template expansion. The BGE description index without expansion still leads on held-out tools (0.764), and the independent pool keeps that ordering (Tab.~\ref{tab:symmetric}).

\paragraph{The two-index router is a retriever-dependent mitigation.} It reserves three of five candidate slots for descriptions and two for experience. The budget was fixed in advance. A later sweep under Jina shows the trade-off it sets: held-out accuracy falls as experience slots are added (by 1.3, 1.5, 4.4 and 10.6 points with one to four), while the gain on labeled queries with no held-out gold tool stays between 3.5 and 4.4 points, so one experience slot does about as well as two (App.~\ref{app:setup}). Over all 729 queries it is the best arm under Jina (0.539 vs.\ 0.528 for \docs) and BM25 (0.584 vs.\ 0.561) and level under BGE (0.602 vs.\ 0.606; Tab.~\ref{tab:twoindex}). Under Jina it recovers most of the held-out accuracy of descriptions (0.520 vs.\ 0.534; McNemar $p=0.11$) but not their recall (0.614 vs.\ 0.689). With the clean model it gains 3.6 held-out points over descriptions (tool-clustered $[+1.4,+5.9]$) and loses 2.9 labeled points ($[-5.1,-0.7]$), and the labeled loss remains on the 340 queries with no held-out gold tool (Tab.~\ref{tab:strata}). We did not test why its held-out accuracy rises while its recall falls (0.821 vs.\ 0.878). Reciprocal-rank fusion halves held-out recall under Jina at a 20-entry pool, and weighted score fusion tuned on held-out development tools sets the experience weight to 0 (App.~\ref{app:setup}). Ranking each tool in the shared index by the mean of its entries, in place of its best entry, shrinks the held-out loss to 10.7 points but gives no labeled gain (App.~\ref{app:setup}).

\paragraph{Secondary: pooled experience beats a client's own on another client's tools.} On the split's pool, each gold tool of the G1-instruction queries gets one owner client that holds all its labeled queries (129 tools, 136 queries), and we score the four other clients. Pooled experience beats each such client's own compendium, 0.605 against 0.425 (+18.0 points, $[+12.2,+24.0]$). Its 6.1-point gain over descriptions has an interval that includes zero, and the test cannot detect gains below about 7 points (Tab.~\ref{tab:e8}).

\section{Using STUN-Bench}

Work that evaluates shared tool experience on the ToolBench pool should do four things. (i) Apply query- and tool-level exclusion before any client or training draw, and release the excluded ids. (ii) Evaluate under a tool-level hold-out and report held-out and labeled queries separately, because exposed tool labels can hide the held-out loss. (iii) Include a lexical classifier fit on the same data, to show what word overlap alone solves. (iv) With general-purpose retrievers, keep pooled experience out of the description index or reserve candidate slots for descriptions, and check labeled accuracy when doing so. When comparing exchange formats, a frozen-candidate harness, gold insertion and a concatenation baseline separate what retrieval proposes from what the reranker reads (App.~\ref{app:controls}).

\section{Conclusion}

STUN-Bench separates query decontamination from tool-level exposure. Removing test-query matches from the ToolBench pool leaves a 5.6-point exposure gap in a trained retriever under matched training. When shared experience excludes the held-out tools, crowding cuts their retrieval from a shared index under general-purpose retrievers and reverses the gain the leaky setup reports. With our clean tool-trained retriever, the end-to-end loss is not detected. The two-index router recovers most of the loss under Jina-v2, but with the clean retriever it trades labeled accuracy for held-out accuracy.

\section*{Limitations}

\paragraph{Exposure scope.} STUN-Bench controls what clients hold and what the clean retriever trains on, not what any base model saw in pretraining. Its conservative mention rule removes 83,114 items (the pool keeps 47\% after both rules) and is worth about a point of held-out recall. The exposure comparisons compare conditions and do not partition causes. Labels and mentions move together even at the matched 0.90 comparison, and optimizer steps differ by about 7.6\% for the 5.2-point step. At 0.90 the size-matched raw model adds no detectable recall over the label-kept one ($+0.5$, $[-0.7,+2.0]$).

\paragraph{Labels and data.} We audited the gold labels of 50 of the 309 held-out queries; the other 259 are unaudited. In the audit, about a fifth of the sampled requests had another tool that served them at least as well as the gold tool. Scoring a route correct only when it hits a gold tool therefore understates absolute accuracy for every arm; the arm comparisons are paired on the same queries. Client experience is simulated: category partitions of the ToolBench pool across five clients, whose queries an LLM wrote from tool descriptions. Real agent logs would be less uniform and less description-like, and the size of crowding under them is untested. The held-out set comes from two test groups (172 tools, 309 queries).

\paragraph{Published systems.} The audit concerns components trained on the pool, retrievers above all. ToolLLM's main generalization experiment gives ToolLLaMA the ground-truth APIs for each instruction \cite{qin2024}, so the audit does not measure it. The raw-pool comparison reproduces how published retrievers draw their data, not their architectures or reported numbers. Our retraining of ToolBench's retriever stays far below the released checkpoint. We do not show that correcting exposure changes any published system's conclusion or ranking.

\paragraph{Evaluation scope.} The full pipeline runs on one catalog and one small agentic benchmark; only the retrieval stage is replicated on further pools. Routing accuracy is top-1 tool selection with any gold tool counted correct. It does not measure whether a route selects every required API, produces valid arguments or completes the task \cite{bhat2026}. End-to-end accuracy is reported for the shared index under Jina-v2 and the clean model; the other retrievers are compared at retrieval. We use one reranker size class (3--8B) plus GPT-4o~\cite{hurst2024gpt4o} on $\tau$-bench, and three to four seeds. The two-index router's slot budget was fixed in advance and swept only under Jina-v2 on the split; on the independent pool it lowers the clean model's held-out recall (0.219 to 0.174). The cross-client test uses one ownership rule and 136 queries.

\section*{Ethics Statement}

The audit reports a property of public benchmark data; we do not claim that any surveyed method is contaminated. Released logs replace query and scenario text with content hashes. We release item ids and code, not query text. A manual review found possibly real personal data (phone numbers, email addresses, postcodes, street addresses) in 2 of 200 sampled pool items and 10 of the 309 held-out queries, all from the public ToolBench release, and no offensive content (App.~\ref{app:datasheet}). Llama-3.1-8B-Instruct generated the expansion queries of Tab.~\ref{tab:symmetric} as part of that experiment. We used generative AI assistants to edit manuscript text, not to generate data, design experiments or run them. The authors take responsibility for all content.

\bibliography{refs}

\appendix

\section{Datasheet for STUN-Bench}
\label{app:datasheet}

\paragraph{Motivation.} We draw client experience from the ToolBench instruction pool, which labels every gold tool of the two G1 ``unseen'' test groups. ToolBench's released retriever is trained on the same pool. The split creates test tools for which no client holds a labeled query. Routing to unlabeled tools can then be measured, and a supervised classifier's ceiling is zero by construction.

\paragraph{Composition.} The split is defined over existing StableToolBench and ToolBench data. It adds the partition and the label-held-out pool. It comprises the 3,180-tool catalog (192 junk entries removed), a client experience pool of 90,704 items (released as item ids with per-item removal reasons), and the 729 junk-filtered StableToolBench evaluation queries, unchanged, in two strata. Held-out tools $H$ are the union of the normalized gold tool ids of the G1-tool (95) and G1-category (77) queries ($|H|{=}172$, released with its content hash). Held-out queries are those whose gold tools are all in $H$ (309). The other 420 are labeled.

\paragraph{Collection and filtering.} From the leak-filtered pool (194,107 items), we remove an item if any label is in $H$ (20,289 items). We also remove it if a normalized held-out tool name appears in its query, gold tool names, or attached tool, API, category name, or API description (83,114 items). 90,704 items remain. Each run asserts that no remaining item labels a tool in $H$, that the evaluation queries are untouched, and that the classifier's held-out accuracy is 0.

\paragraph{Uses and limitations.} The intended use is comparing routing arms on held-out and labeled queries within one run. Labeled-subset numbers are not comparable across pools of different size. The mention rule is conservative and removes 83,114 items. After both rules the pool keeps about 47\%. A category-level construction (App.~\ref{app:setup}, Tab.~\ref{tab:e7}) removes all labels for whole categories and also produces a held-out recall loss at retrieval, smaller than on the split. The split inherits the description-driven nature of StableToolBench queries. It tests routing to unlabeled tools from documentation. The benchmark fixes the gold tools independently of shared experience: we measure how experience about other tools changes routing predictions, not tasks whose correct tool changes with that experience. The cross-client case is a separate construction built on the split (App.~\ref{app:setup}).

\paragraph{Label audit.} We sampled 50 of the 309 held-out queries with a fixed seed, allocated by group (25 G1-tool, 21 G1-category, 4 G2-category; three have two gold tools). Two annotators who are not authors worked independently from written guidelines fixed before annotation. For each query they saw the request, the gold tool descriptions and the description index's top five tools, and no routing outcome. They answered whether the gold tool can serve the request (yes, partly or no, judged from its description; a multi-gold set as a whole) and whether another tool serves it at least as well. Raw agreement before adjudication was 0.90 for the first question (Cohen's $\kappa{=}0.77$) and 0.98 for the second ($\kappa{=}0.94$); the annotators resolved the five disagreements together. After adjudication, 1 of 50 gold labels could not serve its request (Wilson 95\% interval 0.4--10.5\%). Another 23 were incomplete (46\%, 33.0--59.6\%): in 17 the gold tool served only part of the request, which typically asks for several things, and in 10 another catalog tool served it at least as well (4 queries have both). Rescoring the Jina-v2 comparison of Tab.~\ref{tab:heldout} without the 24 flagged queries gives held-out accuracy 0.544 for \docs and 0.296 for the shared index (24.8 points lower, against 25.6 on all 309). Dropping only the incorrect label leaves the loss at 25.6 points. The other 259 held-out queries are unaudited.

\paragraph{Personal data and offensive content.} One reviewer read 200 pool items sampled uniformly with a fixed seed and all 309 held-out queries, following written criteria. Two pool items and ten held-out queries may contain real personal data: phone numbers, email addresses, full postcodes or street addresses tied to a requester, an account identifier, and a named person tied to an employer (Wilson 95\% upper bounds 3.6\% and 5.9\%). No item contained offensive content. All text comes from the public ToolBench release; we release item ids and code, not text.

\paragraph{Distribution and maintenance.} The held-out tool list, per-item removal reasons, subset assignment of the 729 queries, and hashed per-query logs are released with the code at \url{https://abhijit85.github.io/STUN-Bench/} under the license of the source data (ToolBench and StableToolBench are released under Apache-2.0). We do not redistribute the pool. The released filter and per-item removal list reconstruct it from a local copy of the public ToolBench release. The split is versioned with the code, and the tool-list hash identifies it.

\section{Exchange-format controls}
\label{app:controls}

This appendix compares exchange formats on the leak-filtered pool, where all 729 test queries' tools are labeled and no tool is unseen. A new exchange format changes both what retrieval proposes and what the reranker reads, so an accuracy change alone cannot say which helped \cite{zhang2026review}. We apply three controls to our typed merge (\synapse): a frozen-candidate harness, gold insertion, and a no-merge concatenation baseline (\concat, FICAL-style \cite{wu2024fical}: the same client compendiums concatenated with no clustering, majority or conflict handling). The merge has no detected advantage over concatenation.

\begin{table}[t]
\centering\scriptsize
\setlength{\tabcolsep}{3pt}
\begin{tabular}{@{}lcccc@{}}
\toprule
Arm & subset, 0\% & subset, 60\% & full 729 & R@5 \\
\midrule
\synapse & .599\sd{.010} & .581\sd{.029} & .587 & .728 \\
\concat & .593\sd{.007} & .593\sd{.010} & .585\sd{.007} & .724 \\
\bottomrule
\end{tabular}
\caption{Typed merging and plain concatenation are equivalent within a 2-point margin on all 729 queries at the 0.95 filter; at 60\% contradiction no difference is detected but equivalence is not shown. \concat: no clustering, majority, or conflict handling. Subset: G1+G2-instruction ($n{=}258$) at 0\% and 60\% injected contradiction.}
\label{tab:merge}
\end{table}

\paragraph{Typed merging and concatenation are equivalent within 2 points at the 0.95 filter.} \concat scores 0.585 and \synapse 0.587 on the full benchmark (recall@5 0.724 vs.\ 0.728; Tab.~\ref{tab:merge}). The query-clustered 95\% interval on the difference, $[-0.55,+0.82]$ points, lies inside a margin of 2 points declared before the run. The same holds on the G1+G2-instruction subset at 0\% injected contradiction. At 60\% the interval crosses the margin, and no difference is detected. At the 0.90 filter (one seed), \synapse is 5.8 points below \concat, so the equivalence holds at 0.95 only. On byte-identical candidate lists, rendering typed fields instead of one JSON string has no detected effect under two model families (Tab.~\ref{tab:rendering}); this does not show equivalence.

\paragraph{No federation-related accuracy difference is detected.} \synapse is within 1.1 points of \central on every group (Tab.~\ref{tab:groups}; 0.587 vs.\ 0.585 overall, McNemar $p=0.79$). Experience adds 6--11 points over \docs on single-tool queries and 1--4 points on multi-tool ones; this gain does not survive the 0.90 filter (Tab.~\ref{tab:threshold}).

\begin{table}[t]
\centering\scriptsize
\setlength{\tabcolsep}{2.4pt}
\begin{tabular}{@{}lccccccc@{}}
\toprule
 & \multicolumn{3}{c}{G1} & \multicolumn{2}{c}{G2} & G3 & \\
\cmidrule(lr){2-4}\cmidrule(lr){5-6}\cmidrule(lr){7-7}
Arm & instr. & tool$^{\dagger}$ & cat.$^{\dagger}$ & instr. & cat.$^{\dagger}$ & instr. & all \\
 & (153) & (152) & (134) & (105) & (124) & (61) & (729) \\
\midrule
\local & .501 & .479 & .496 & .322 & .306 & .283 & .418 \\
\docs & .562 & .546 & .537 & .533 & .460 & .508 & .528 \\
\synapse & .619 & .627 & .649 & .571 & .479 & .514 & .587 \\
\central & .614 & .616 & .652 & .562 & .489 & .525 & .585 \\
\clf & .837 & .827 & .838 & .816 & .780 & .771 & .817 \\
\bottomrule
\end{tabular}
\caption{With every test tool labeled, the lexical classifier leads every LLM arm in every group, including the ``unseen'' ones. Routing accuracy on the leak-filtered pool, three seeds; $K{=}5$ clients, 5,000 items each. $^{\dagger}$Called unseen by ToolBench; the raw pool labels these tools (\S\ref{sec:audit}).}
\label{tab:groups}
\end{table}

\paragraph{Contradiction injection.} Before the merge we inject scenario/precaution pairs about real tools (``use tool $t$ for request $x$'' vs.\ ``do not'') at rates $\{0,0.2,0.4,0.6\}$ of client contributions. The injected set is identical across arms for each (rate, seed) and verified by hash. Seeds are 7/42/456/1234, except that the round-delayed policy and the 20\% column use three of them.

\paragraph{Whole-string control.} \flatm serializes each scenario's fields into one string and takes a whole-string majority. In its runs the reranker's fallback path rendered only the first 200 characters of each candidate (App.~\ref{app:setup}). We therefore base no claim on it and use \concat as the untyped control. At frozen candidates, content is rendered in full. There, the \flatm gap to the typed merge (4.7 and 2.0 points) fits the difference between field-wise and whole-string voting. The gap disappears against concatenation, which does not vote.

\begin{table*}[t]
\centering\small
\setlength{\tabcolsep}{2.6pt}
\begin{tabular}{@{}llcccc@{}}
\toprule
 & & \multicolumn{2}{c}{Llama-3.1-8B} & \multicolumn{2}{c}{Qwen2.5-7B} \\
\cmidrule(lr){3-4}\cmidrule(lr){5-6}
Merge & Rendered as & 0\% & 60\% & 0\% & 60\% \\
\midrule
typed & typed fields & .583\sd{.018} & .575\sd{.026} & .588\sd{.030} & .581\sd{.014} \\
typed & one JSON string & .581\sd{.014} & .559\sd{.016} & .580\sd{.022} & .558\sd{.021} \\
flat & decoded fields & .497\sd{.026} & .501\sd{.018} & .522\sd{.011} & .539\sd{.015} \\
flat & one JSON string & .534\sd{.023} & .539\sd{.012} & .549\sd{.014} & .555\sd{.009} \\
\midrule
\multicolumn{2}{@{}l}{R@5 of frozen lists, typed / flat} & .764 / .811 & .769 / .822 & \multicolumn{2}{c}{same lists} \\
\bottomrule
\end{tabular}
\caption{For the typed merge, rendering fields or one JSON string makes no detected difference. Merge crossed with rendering on frozen candidate lists (G1+G2-instruction, $n{=}258$; seeds 42/123/456). Both models rerank byte-identical lists, so recall@5 is a property of the merge. Qwen2.5-3B (App.~\ref{app:setup}) shows the same pattern.}
\label{tab:rendering}
\end{table*}

\begin{table}[t]
\centering\footnotesize
\setlength{\tabcolsep}{4pt}
\begin{tabular}{@{}lccc@{}}
\toprule
Arm & retr.\ top-1 & routed & gold ins. \\
\midrule
\synapse & .476 & .587 & .612\sd{.012} \\
\central & .478 & .585 & .613\sd{.010} \\
\docs & .422 & .528 & .658 \\
\clf & -- & .817 & (no retrieval) \\
\bottomrule
\end{tabular}
\caption{Forcing the gold tool into the candidates lifts descriptions by 13 points and the compendium by only 2.5. All 729 queries, three seeds. Retrieval top-1: highest-ranked distinct tool, no reranker. Gold ins.: the gold tool replaces the lowest-ranked non-gold candidate, then the same reranker runs.}
\label{tab:loss}
\end{table}

\begin{figure}[t]
\centering
\includegraphics[width=\columnwidth]{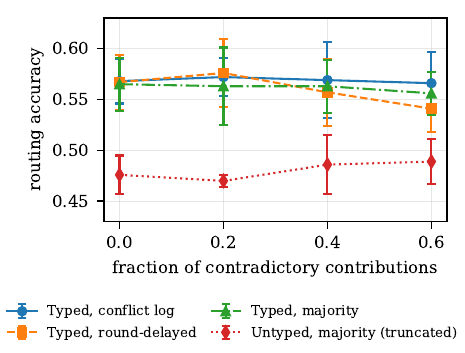}
\caption{Injected contradiction on G1+G2-instruction ($n{=}258$). No difference among the three typed policies is detected at any rate. \flatm's candidates were rendered as their first 200 characters, so we draw no claim from it.}
\label{fig:contradiction}
\end{figure}

\paragraph{Rendering the typed merge: no difference detected on frozen candidates.} Table~\ref{tab:rendering} crosses merge and rendering on frozen lists. For the typed merge, decoded fields and one JSON string in the same labeled slots give 0.583 vs.\ 0.581 at 0\% and 0.575 vs.\ 0.559 at 60\% under Llama-3.1-8B. Qwen2.5-7B gives 0.588 vs.\ 0.580 and 0.581 vs.\ 0.558 on the same lists, and Qwen2.5-3B behaves alike. Cells with a nominal gap are within about one SD. These are null results. At 258 queries per cell the query-clustered intervals admit differences of up to 2.6 points, so equivalence is not shown. For the flat merge the sign of the rendering effect depends on the model. Decoded fields score below one JSON string under Llama-3.1-8B and Qwen2.5-7B, and above it under Qwen2.5-3B. Both serializations keep the field labels, so this tests rendering alone. Whether structure matters is a separate question.

\paragraph{Gold insertion on the leak-filtered pool.} Table~\ref{tab:loss} compares retrieval-only top-1, real routing, and gold insertion on all 729 queries. Forcing every gold tool into the five candidates lifts \docs from 0.528 to 0.658, but it lifts \synapse only from 0.587 to 0.612\sd{.012}. Perfect recall is thus worth about 13 points to descriptions and 2.5 points to the compendium. With recall equalised, the compendium ends below descriptions alone, although the two arms still differ in their four non-gold candidates. Experience earns its labeled-tool advantage at the retrieval stage, where it raises recall@5 from 0.672 to 0.728. The smaller lift under gold insertion fits part of that advantage being given back at ranking. But the arms' distractors differ, so we cannot attribute it to ranking alone. Retrieval misses account for 13 of the 29 points between \docs and the classifier, a substantial part but less than half.

\paragraph{Conflict handling: not measurable on this benchmark.} Figure~\ref{fig:contradiction} injects contradiction at up to 60\%. No difference among the three typed policies is detected at any rate (query-clustered intervals up to 2.2 points). Contradiction costs any arm at most about 3 points. On StableToolBench a contradiction never changes the gold tool, because queries are generated from descriptions \cite{qin2024}. No experiment here can therefore measure the mechanism that motivated typing.

\section{Released artifacts}
\label{app:artifacts}

Code, the leak filter, STUN-Bench, and per-query logs are available at \url{https://abhijit85.github.io/STUN-Bench/}. The released code produces every artifact from public inputs, and the release includes the scripts that regenerate each table.

\begin{itemize}\setlength\itemsep{0pt}
\item Leak filter. Input: ToolBench instruction pool, StableToolBench test queries, Jina-v2 embeddings. Output: pool with exact and cos $\ge 0.95$ duplicates removed (1,916 items); report of cos $\ge 0.90$ near-duplicates (319/729); per-item removal reasons.
\item STUN-Bench (label-held-out split). Input: filtered pool, G1-tool and G1-category groups. Output: held-out tool list $H$ (172, hashed); pool with labeled and mentioned items removed (90,704 of 194,107); subset assignment for all 729 queries.
\item Frozen-candidate harness. Input: a merged artifact, an index builder, a reranker. Output: per-query top-20 lists; recall assertion against the source cell; byte-identity assertion across renderings; paired per-query outputs.
\item Gold-insertion intervention. Same plus gold tools. Output: accuracy with gold tools inserted into the top-5; insertion fraction per query.
\item Concatenation control. Input: client artifacts. Output: edge artifact with no clustering, majority, or conflict handling, rendered through the same prompt.
\item Hashed per-query logs. Every run: query and scenario text replaced by content hashes; candidate ids, prediction, correctness, code revision, and hashes of loaded artifacts.
\end{itemize}

\section{Federation paradigms compared}
\label{app:paradigms}

\begin{table*}[t]
\centering\scriptsize
\setlength{\tabcolsep}{3pt}
\begin{tabular}{@{}lllll@{}}
\toprule
Paradigm (examples) & Exchanged unit & Frozen, heterog. & Privacy unit & Conflict \\
\midrule
Weight / adapter FL \cite{ye2024,wu2024fedbiot} & parameters & no & gradient / adapter & averaged away \\
Distillation FL \cite{li2019fedmd,lin2020} & logits on a shared set & yes, with shared data & prediction vector & averaged away \\
Federated RAG \cite{zhao2024frag,addison2024} & passages / embeddings & yes & passage & none (both retrieved) \\
Federated prompts \cite{chen2025fedtextgrad,wang2025fedicl} & prompt text / examples & yes & whole text & none or overwritten \\
Concat.\ compendium (FICAL \cite{wu2024fical}; \concat) & named parts, concatenated & yes & whole text & none (both kept) \\
Typed compendium (\synapse) & schema-validated fields & yes & per field & field majority + dissent log \\
\bottomrule
\end{tabular}
\caption{What each federation paradigm exchanges and what its merge can therefore check. The last row is the typed unit under test. Under controls, App.~\ref{app:controls} detects no accuracy advantage over the concatenated compendium (\concat); the other rows were not compared experimentally.}
\label{tab:paradigms}
\end{table*}

Table~\ref{tab:paradigms} summarizes, for each family in \S2, what object crosses the boundary and what its merge can check.

\section{Reuse of the ToolBench instruction pool in published methods}
\label{app:reuse}

The audit concerns the raw ToolBench instruction files. We draw client data from them, and ToolBench's released retriever is trained on them. The methods of Tab.~\ref{tab:reuse} learn from ToolBench data. Two facts about the official release set how far the audit reaches. First, the released exclusion is by query id only and covers one test set. The released retriever preprocessing excludes the query ids of one list, and its documented example passes the single-tool instruction test list with the single-tool query file. It does no tool- or category-level exclusion. The released preprocessing for ToolLLaMA's fine-tuning data has no test-id exclusion option. ToolLLaMA's released train and eval files are pre-split, and the script cannot reconstruct that split. The ToolBench paper defines the Tool and Category test groups as ``unseen tools'' without describing a procedure that removes them \cite{qin2024}. Table~\ref{tab:lineage} traces each data source. Second, at least one independent paper reports having to do that removal itself. ToolRerank removes tools and categories from ToolBench's training data so that the APIs of the I1-Tool, I1-Cat and I2-Cat queries are unseen in training \cite{zheng2024}. LoSemB constructs its own inductive splits \cite{zhuang2026}. To within about a hundred queries, the released retrieval training set is the whole G1 pool over the whole G1 tool corpus. It has 87,322 unique training queries against 87,419 pool queries (from a public issue on the ToolBench repository and \citet{chen2024reinvoke}, Table~5).

\begin{table*}[t]
\centering\scriptsize
\setlength{\tabcolsep}{3pt}
\begin{tabular}{@{}p{0.27\textwidth}p{0.38\textwidth}p{0.29\textwidth}@{}}
\toprule
Source & Exclusion rule & Exact overlap with StableToolBench test queries \\
\midrule
Raw ToolBench instruction pool & none & 765/765 raw test queries (783 pool rows) \\
Official ToolBench retriever example & single-tool query file minus the single-tool instruction test ids only & 314/765 remain (3 G1-instruction duplicates, 158 G1-tool, 153 G1-category) \\
ToolLLaMA fine-tuning preprocessing & no test-id exclusion; released train file is pre-split & not auditable from the script alone \\
Our post-junk experience pool & junk-tool filter only & 729/729 post-junk queries; 196,023 items \\
Our leak-filtered pool & all six evaluation groups removed: exact matches and Jina cosine $\ge 0.95$ & 745 exact and 1,171 near-duplicate rows removed; 194,107 items \\
Our label-held-out pool & leak-filtered pool minus held-out-tool labels and mentions & 90,704 items; no held-out labels \\
Our label-kept pool & leak-filtered pool sampled uniformly to 90,704, held-out-tool items not excluded & 0 exact; 9,482 held-out-labeled items (mean over seeds) \\
\bottomrule
\end{tabular}
\caption{Only tool-level exclusion removes the held-out tools. The official retriever example still leaves 314 of 765 test queries in the pool. Data lineage: which exclusion each source applies and how much test-query overlap remains.}
\label{tab:lineage}
\end{table*}

Table~\ref{tab:reuse} lists the methods we found (searched September 2026) that learn a model, retriever, or memory from ToolBench data and evaluate on ToolBench or StableToolBench test splits. For each, it records what the paper states about exclusion. The columns record what the paper says, not what its code does. ``Not stated'' means the paper contains no statement. Several of these methods may well have excluded test queries without saying so. Of sixteen learning methods, two state that they excluded the test queries, and two state that they excluded the held-out tools. The tool-retrieval works draw from the pools without tool-level exclusion. COLT samples its own split from the raw I2/I3 files. The iterative-feedback retriever trains on 195,937 instructions (the three pools less the test queries). ToolGen uses 194k I1 pairs for retrieval and covers every API in the corpus in its memorization stage. ToolGen's own appendix discusses this stage-wise coverage \cite{wang2025toolgen}. So ``not stated'' records the absence of a stated exclusion, not unawareness of exposure. Documentation exposure during memorization also differs from the labeled usage exposure we audit. We found no prior report that the instruction pool contains the StableToolBench test queries verbatim or labels every gold tool of the G1 ``unseen'' groups.

\paragraph{One method's released code.} Tool-MVR \cite{ma2025} reports StableToolBench results on all six test groups, with the ``unseen'' G1 groups level with the instruction group (G1-Tool 86.7 and G1-Category 83.0 against G1-Instruction 83.4 for the Qwen variant). Its paper does not state an exclusion. Its released code builds training data from the raw G1, G2 and G3 query files. Before fine-tuning, it drops trajectories whose query id is in any of the six StableToolBench test files. That is a query-id exclusion with no tool-level step, so the source it filters still holds labeled queries for every held-out tool (\S\ref{sec:audit}). Only three examples of each released training set are public, so we cannot count how many of those queries survive its quality filter. We make no claim about its reported numbers.

\begin{table}[t]
\centering\scriptsize
\setlength{\tabcolsep}{2.5pt}
\begin{tabular}{@{}lccc@{}}
\toprule
Training data & G1-tool$^{\dagger}$ & G1-cat.$^{\dagger}$ & G1-instr. \\
\midrule
A: released training set & .696\sd{.034} & .777\sd{.019} & .721\sd{.011} \\
B: held-out tools removed & .330\sd{.041} & .307\sd{.043} & .739\sd{.026} \\
C: size-matched raw sample & .705\sd{.023} & .827\sd{.026} & .695\sd{.021} \\
\midrule
Released checkpoint & .953 & .982 & .981 \\
\bottomrule
\end{tabular}
\caption{Removing the held-out tools' items cuts recall on the unseen groups. A size-matched raw sample does not. ToolBench's retriever retrained on three training sets, recall@5 on ToolBench's official retrieval evaluation, three seeds. $^{\dagger}$Called unseen.}
\label{tab:tbir}
\end{table}

\paragraph{ToolBench's retriever, retrained.} We reran the training recipe of ToolBench's released retriever (ToolBench-IR) on three training sets, with three seeds each. We scored every checkpoint with ToolBench's official retrieval evaluation (Tab.~\ref{tab:tbir}). Arm A uses the released training set (197,873 pairs, 30,920 updates). Arm B removes the held-out tools' items (175,637 pairs, 27,445 updates). Arm C is a sample of the raw training data matched to B's size (176,534--176,740 pairs, 27,585--27,620 updates). On the two ``unseen'' groups, B falls far below A: G1-tool recall@5 is 0.330 against 0.696, and G1-category recall@5 is 0.307 against 0.777. Both gaps exceed eight seed SDs. C does not fall (0.705 and 0.827). On G1-instruction, B is level with A (0.739 against 0.721). We read this as evidence that the drop follows the held-out tools' items and not the smaller training set. We have not run query-level paired tests for this comparison. Our arm A does not reproduce the released checkpoint, which scores 0.953, 0.982, and 0.981 on the three groups. The gap is at least as large on G1-instruction as on the ``unseen'' groups, so it is a difference of recipe or checkpoint and not of exposure. We did not find its cause. We therefore compare A, B, and C only with one another.

\paragraph{Propagation to derived corpora.} ToolRet \cite{shi2025toolret} releases a training corpus, ToolRet-train, of 208,826 query--tool pairs assembled from ToolACE, APIGen, and ToolBench. The corpus has no per-row source field. In three of its dense categories (Tools, Music, News Media), 1000/1000 sampled queries match a ToolBench instruction-pool query exactly. Against the 729 StableToolBench evaluation queries, 532/729 appear verbatim and 537/729 at cosine $\ge 0.95$, so a tool retriever trained on this corpus has seen most of the test set. We therefore do not use ToolRet-train as an independent benchmark. The independent pool draws only on the evaluation sets of the ToolRet benchmark. We did not audit APIGen, whose executable APIs come from ToolBench, and make no claim about retrievers trained on it. The ToolRet benchmark's sparse evaluation configurations carry their source dataset in the configuration name. We dropped those named for ToolBench or APIGen. We then checked the remaining 32 configurations (5,664 queries) against the ToolBench instruction pool and found 0 exact matches and 4 near-duplicates. That pool is the independent block of Tab.~\ref{tab:retrievers}. Tools with at least two queries are eligible, and each seed holds out 30\% of them at random (350 tools). Every other eligible tool contributes one query as a scenario. Held-out tools contribute none.

\begin{table*}[t]
\centering\scriptsize
\setlength{\tabcolsep}{3.5pt}
\begin{tabular}{@{}lllllll@{}}
\toprule
Method & Venue & Learns & Draws from & Q-excl. & T-excl. \\
\midrule
ToolLLaMA$^{\ddagger}$ \cite{qin2024} & ICLR'24 & agent + retriever & official G123 train & one id list (retriever) & not stated \\
ToolGen \cite{wang2025toolgen} & ICLR'25 & memorize + retrieve + agent & full pools; all 47k APIs & not stated & not stated \\
COLT \cite{qu2024} & CIKM'24 & tool retriever & raw I2/I3, own 90/10 split & own & not stated \\
Re-Invoke baseline \cite{chen2024reinvoke} & EMNLP-F'24 & ToolLLM retriever & full pools (eval on pools) & n.a. & n.a. \\
ProTIP \cite{anantha2023} & arXiv'23 & retriever + planner & ToolBench train/test (cleaned) & existing split & not stated \\
ToolRet-train \cite{shi2025toolret} & ACL-F'25 & tool retrievers & ToolBench training sets & not stated & not stated \\
StepTool \cite{yu2025} & CIKM'25 & SFT + RL & ToolBench training tasks & not stated & not stated \\
TP-LLaMA \cite{chen2024tpllama} & NeurIPS'24 & SFT + DPO & ToolBench trajectories & yes (by instruction) & not stated \\
$\alpha$-UMi \cite{shen2024} & EMNLP'24 & multi-LLM agent & 62,694 sampled trajectories & not stated & not stated \\
ToolPlanner \cite{wu2024toolplanner} & EMNLP'24 & SFT + RL & G3 seed tasks & yes & n.a. \\
ToolRerank \cite{zheng2024} & LREC-COLING'24 & reranker & official train, tools removed & not stated & yes \\
DTA-Llama \cite{zhu2025dta} & ACL'25 & SFT & ``training portion'' & implicit & not stated \\
Tool-MVR$^{\ddagger}$ \cite{ma2025} & KDD'25 & SFT + reflection & re-verified G1/G2/G3 queries & not stated & not stated \\
AutoTools \cite{shi2025wild} & WWW'25 & understanding + trajectories & ToolBench among others & not stated & not stated \\
Tool2Vec \cite{moon2024} & arXiv'24 & fine-tuned encoders & I1/I2/I3 (original codebase) & not stated & not stated \\
Iterative feedback \cite{xu2024} & EMNLP-F'24 & retriever & 195,937 training instructions & not stated & not stated \\
LoSemB \cite{zhuang2026} & WWW'26 & inductive retriever & own inductive splits & own & yes (own) \\
\bottomrule
\end{tabular}
\caption{Published methods that learn from ToolBench data and evaluate on ToolBench or StableToolBench. Q-/T-excl.: whether the paper states that test queries / held-out tools were excluded (statements, not code). Own: the paper's own split. n.a.: no held-out-tool group. $^{\ddagger}$Code inspected (App.~\ref{app:reuse}). For ToolLLaMA, the retriever recipe was also retrained (Tab.~\ref{tab:tbir}).}
\label{tab:reuse}
\end{table*}

\section{Experimental setup and provenance}
\label{app:setup}

\paragraph{Guide to the appendices.} This appendix gives the setup and the statistical procedures. Its tables list the protocol departures (Tab.~\ref{tab:exceptions}), training-pool sizes (Tab.~\ref{tab:pairs}), mixed-query results (Tab.~\ref{tab:strata}), and diagnostic arms (Tab.~\ref{tab:diag}). App.~\ref{app:controls} reports the exchange-format controls, App.~\ref{app:datasheet} specifies STUN-Bench, and App.~\ref{app:reuse} gives the data lineage and the retraining of ToolBench's retriever.

\paragraph{StableToolBench.} The registry holds 3,180 tools from the maintained release \cite{guo2024} after we remove 192 junk entries. A junk entry has an empty or missing description, a normalized name that matches the token ``test'', ``demo'', ``asdf'' or ``placeholder'', or a name with no ASCII letter or digit. We then drop an evaluation query or pool item only if all of its gold tools were removed. This takes the test set from 765 to 729 queries. Client experience comes from the ToolBench instruction pool, leak-filtered as in \S\ref{sec:audit}. Each client gets 5,000 items, partitioned by category (non-IID). The test set is the six public groups of solvable queries (those a majority of three LLM judges rated solvable \cite{guo2024}), 729 queries after the junk filter. In ToolBench's naming, instruction groups pose new queries over training tools. The tool and category groups use tools ToolBench calls unseen, from seen and unseen categories respectively \cite{qin2024}. Gold is a set of tool ids. We normalize ids by trimming, lowercasing, and replacing whitespace with underscores. Candidate labels get the same normalization. A route is correct if the predicted tool is in the gold set. Recall@5 counts a hit if any gold tool is among the five candidates. Each client builds its precautions from observed near-misses. A near-miss is a query whose top-cosine candidate was a tool outside the gold set.

\paragraph{Retrievers.} All retrievers run locally. Jina-v2: Jina Embeddings v2 base (English), batch 64, L2-normalized. BGE: BGE-base-en v1.5. BM25: Okapi on alphanumeric word tokens, $k_1{=}1.2$, $b{=}0.75$. Qwen3-Embedding-8B, with the fixed query instruction ``\texttt{Instruct: Given a user request, retrieve the tool that can fulfil it\textbackslash nQuery: }'' prepended to queries only. Tool descriptions and scenarios are encoded without it. We fixed the instruction before the run and did not tune it. Without it, Qwen3's held-out recall on the independent pool falls further (0.099 from descriptions, 0.082 from the shared index, against 0.117 and 0.102). The instruction therefore does not cause its low score there. Under the shared index, labeled recall rises slightly for Qwen3 (0.729 to 0.749) and for ToolRet-e5 (0.857 to 0.885). In Tab.~\ref{tab:gap}, the gold score is the maximum over a query's gold tools. Seed SDs of the mean gap are at most 0.002 for the dense retrievers and 0.48 for BM25.

\paragraph{Routing constants.} Retrieval ranks compendium units by the cosine similarity of their Jina-v2 embeddings and keeps the five highest-ranked distinct tools. The reranker scores these with the description-first prompt. To collect those five, a routing run walks down the ranking until it has five distinct tools (cap 200 entries). The contradiction and frozen-candidate runs are the exception: they filter a fixed pool of 20 entries to distinct tools. The edge aggregator clusters scenarios at cosine $\ge 0.85$. The leak filter removes exact matches and near-duplicates at cosine $\ge 0.95$. It reports the remaining matches at cosine $\ge 0.90$ without removing them. Each run records the hash of its prompt template.

\paragraph{Router prompt.} The prompt shows each candidate tool's description first, then its scenarios (``when to use'') and merged precautions plus rendered conflict log (``do not use when''). The reranker scores candidate indices by next-token log-probability. In the frozen-candidate protocol the same slots hold either decoded fields or the candidate's content as one JSON string. The labels are printed in both cases.

\paragraph{Whole-string control rendering.} The \flatm arm stores only its serialized payload. When a candidate has no typed fields, the reranker's fallback path renders the first 200 characters of the candidate's raw text. The reranker therefore sees only the opening fragment of that arm's candidates, and we base no claim on it (App.~\ref{app:controls}). The frozen-candidate runs render the full content.

\paragraph{Frozen-candidate runs.} Each run loads a merged compendium from a completed contradiction cell and checks it against the recorded content hash. It builds the index once per merge cell and requires its recall@5 to equal the original cell's to three decimals. The candidate lists go to disk. We rerank every rendering on those same lists. The run asserts that the lists are byte-identical per query. Llama-3.1-8B-Instruct, Qwen2.5-7B-Instruct, and Qwen2.5-3B-Instruct all use these lists. Qwen2.5-3B scores: typed/typed-fields .554\sd{.016} (0\%), .542\sd{.030} (60\%); typed/JSON .556\sd{.018}, .530\sd{.016}; flat/decoded .535\sd{.023}, .535\sd{.007}; flat/JSON .516\sd{.028}, .527\sd{.021}.

\begin{table}[t]
\centering\scriptsize
\setlength{\tabcolsep}{2pt}
\begin{tabular}{@{}lcccc@{}}
\toprule
 & \multicolumn{2}{c}{held-out (309)} & \multicolumn{2}{c}{labeled (420)}  \\
\cmidrule(lr){2-3}\cmidrule(lr){4-5}
Arm & acc & R@5 & acc & R@5  \\
\midrule
Separate index$^{a}$ & 0.531 & 0.686 & -- & --  \\
Two-index, RRF & 0.234\sd{.018} & 0.331\sd{.009} & 0.511\sd{.027} & 0.673\sd{.010}  \\
\docs, gold ins. & 0.644 & ins.\ 0.343 & -- & --  \\
\synapse, gold ins.$^{b}$ & 0.526\sd{.002} & ins.\ 0.699\sd{.006} & -- & --  \\
\bottomrule
\end{tabular}
\caption{Diagnostic arms on STUN-Bench under Jina, mean$\pm$SD over seeds 42/123/456. Gold insertion lifts both arms but leaves the shared index below descriptions. Rank fusion loses held-out tools. $^{a}$Separate index: candidates come from the description index only and are reranked with the compendium rendering. Experience never enters retrieval, unlike in the two-index router. RRF: reciprocal-rank fusion of the two indices. Gold ins.: the gold tool replaces the lowest-ranked non-gold candidate. Ins.: share of queries that needed it. $^{b}$Separate compendium of the split.}
\label{tab:diag}
\end{table}

\paragraph{Held-out split.} Held-out tools $H$ are the union of gold tools of the G1-tool and G1-category test queries, selected after the junk filter (172 tools, released with a content hash). A query is held out only if all of its gold tools are in $H$. On the ToolBench single-tool pool, each seed instead holds out 30\% of the eligible G1 tools at random (482 tools). We remove a pool item if any label is in $H$. We also remove it if a normalized held-out tool name appears in its query, its gold tool names, or the tool name, API name, category name, or API description attached to it. Normalization lowercases names and maps underscores and hyphens to spaces. The separate-index run of Tab.~\ref{tab:diag} evaluated the 309 held-out queries and an 80-query labeled subset against a description index rebuilt for that run. We therefore report only its held-out cells. The two-index rows and Tab.~\ref{tab:retrievers} rebuild the compendium and both indices per seed and evaluate all 729 queries. That is why their \docs cells can differ from Tab.~\ref{tab:heldout} in the third decimal. \docs has no experience input. Its scores are identical across filter modes to six decimals. Tables report mean$\pm$SD over seeds and omit SDs of .000. A cell without an SD has no seed-dependent input. In Tab.~\ref{tab:heldout}, seed 123 of \local{} stopped on an underfilled list (below). Tab.~\ref{tab:retrievers} comes from a separate run, so its Jina shared cell differs from Tab.~\ref{tab:heldout}'s.

\begin{table}[t]
\centering\scriptsize
\setlength{\tabcolsep}{2pt}
\begin{tabular}{@{}p{0.36\columnwidth}p{0.6\columnwidth}@{}}
\toprule
Run & Departure from the default protocol \\
\midrule
\local, split & seeds 42/456; seed 123 stopped on an underfilled list \\
Held-out gold insertion & persisted retriever-comparison compendium of the same split \\
Separate-index row & 309 held-out and an 80-query labeled subset; held-out cells only \\
Contradiction grid, \flatm & fixed-depth candidate lists; \flatm rendered as 200 characters \\
Contradiction seeds & 7/42/456/1234; round-delayed policy and 20\% column use three \\
0.90 filter, \synapse, \central & seed 42 only \\
0.90 filter, \concat & walk-down cap 5,000 entries (default 200) \\
$\tau$-bench & two harness revisions; \docs replicated on the later one \\
\bottomrule
\end{tabular}
\caption{Runs that depart from the default protocol of \S\ref{sec:protocol} (three seeds, five-distinct-tool walk-down capped at 200 entries, full rendering). Each departure is explained where the run is reported.}
\label{tab:exceptions}
\end{table}

\paragraph{Departures from the default protocol.} Table~\ref{tab:exceptions} collects every run that departs from the default seeds, candidate rule, or rendering. Each is explained where it is reported.

\paragraph{Candidate budget and underfilling.} Every reported number outside the contradiction grid and the frozen-candidate runs uses the walk-down rule. A fixed number of entries would underfill an index that holds several entries per tool. Retrieving five entries and de-duplicating by tool leaves the shared index with 2.67 distinct candidates on average (fewer than five on 0.920 of queries). That rule would report 0.248 held-out accuracy and 0.275 recall for \synapse, against 0.278\sd{.006} and 0.333\sd{.006} under the walk-down rule. Underfilling would thus add about 3 accuracy and 6 recall points to the loss. On the ToolBench single-tool and independent pools a depth of 20 filled five distinct tools on every query. The description index's held-out recall is 213/309 under both Jina and BM25, on different query sets. We report \local over seeds 42 and 456. On seed 123, one client's index holds too few tools to supply five distinct candidates for one query. The run stopped instead of reporting an underfilled list. Tabs.~\ref{tab:groups}, \ref{tab:merge}, and \ref{tab:threshold} rerank merged artifacts loaded by content hash under the same rule. Under gold insertion the rule matters in the other direction. With two or three distinct candidates, the experience arms would score about 15 points higher (\synapse 0.762 against 0.612\sd{.012}). \docs is unchanged at 0.658. The held-out gold-insertion rows of Tab.~\ref{tab:diag} come from the persisted retriever-comparison compendium of the same split. The run behind the accuracy rows did not keep its merged artifacts. The two builds agree on the one quantity both report, shared-index held-out recall@5 (0.333\sd{.006} against 0.335\sd{.007}). The contradiction grid of Fig.~\ref{fig:contradiction} and the \flatm arm use fixed-depth lists. Every arm there shares that rule, and we compare arms only within the figure. The frozen-candidate runs are unaffected because the renderings are scored on byte-identical lists. All runs are in the released records with their candidate rule.

\begin{table}[t]
\centering\scriptsize
\setlength{\tabcolsep}{3pt}
\begin{tabular}{@{}llccc@{}}
\toprule
Retriever & Arm & held-out & labeled & all \\
\midrule
Jina-v2 & \docs & .534 & .524 & .528 \\
Jina-v2 & shared index & .278\sd{.006} & .528\sd{.009} & .422\sd{.003} \\
Jina-v2 & two-index router & .520\sd{.002} & .553\sd{.015} & .539\sd{.008} \\
Jina-v2 & two-index, RRF & .234\sd{.018} & .511\sd{.027} & .394\sd{.014} \\
BM25 & \docs & .576 & .550 & .561 \\
BM25 & two-index router & .538\sd{.007} & .618\sd{.012} & .584\sd{.010} \\
BGE & \docs & .595 & .614 & .606 \\
BGE & two-index router & .579\sd{.006} & .618\sd{.006} & .602\sd{.002} \\
\bottomrule
\end{tabular}
\caption{Two-index router by retriever on the split, three seeds. Two-index router: three description and two experience slots. RRF: reciprocal-rank fusion of the two rankings.}
\label{tab:twoindex}
\end{table}

\paragraph{Two-index router.} Table~\ref{tab:twoindex} gives the router under each retriever. We fixed the budget of three description and two experience slots in advance. The sweep below came later and did not change it. The router takes three description candidates, fills to five from the experience index, and backfills from descriptions if experience offers no new tool. The two lists are de-duplicated by tool. The same prompt as in every other arm then reranks the five candidates.

\paragraph{Fusion rules.} Recomputing the three-plus-two slot budget reproduces the two-index router of Tab.~\ref{tab:twoindex} under every retriever (0.539\sd{.005}, 0.585\sd{.008}, 0.602\sd{.003} against 0.539\sd{.008}, 0.584\sd{.010}, 0.602\sd{.002}). Reciprocal-rank fusion uses $k{=}60$. We first deduplicate each source ranking (description index, experience index) to distinct tools and keep each tool's highest-scoring entry. Then we assign ranks. A tool scores $\sum 1/(k+\mathrm{rank})$ over the lists it appears in, and a list it is absent from contributes nothing. Additive reciprocal-rank fusion over a 200-entry pool drives held-out recall to 0.000 under all three retrievers. At the 20-entry pool of Tab.~\ref{tab:diag}, it halves held-out recall. Across all seeds and queries, fusion never surfaced a held-out gold tool that the description list had missed. Weighted score fusion standardises each index's scores and combines them with one weight, chosen on a development split. We divide the held-out tools into development and test tools. In seed 42 this gives 146 development queries, 153 test queries, and 10 queries that span both (309 in all). We choose the weight on the development queries and evaluate it on the test queries. The chosen weight was 0 for every retriever and seed. The rule therefore reduces to the description index. On the test queries it scores 0.484, 0.536, and 0.549 (recall@5 0.647, 0.686, 0.719). We did not test an objective that also weights labeled and cross-client queries, which might give experience a nonzero weight.

\begin{table*}[t]
\centering\scriptsize
\setlength{\tabcolsep}{5pt}
\begin{tabular}{@{}lcccccc@{}}
\toprule
 & \multicolumn{2}{c}{held-out} & \multicolumn{2}{c}{labeled} & \multicolumn{2}{c}{mixed} \\
Arm & Acc & R@5 & Acc & R@5 & Acc & R@5 \\
\midrule
5+0 (\docs) & .534 & .689 & .512 & .641 & .575 & .738 \\
4+1 & .521\sd{.003} & .650 & .556\sd{.012} & .723\sd{.010} & .588\sd{.013} & .721\sd{.014} \\
3+2 (router) & .519\sd{.004} & .612 & .552\sd{.012} & .730\sd{.012} & .562\sd{.012} & .696\sd{.014} \\
2+3 & .490\sd{.004} & .560 & .547\sd{.003} & .726\sd{.010} & .512\sd{.025} & .621\sd{.019} \\
1+4 & .428\sd{.005} & .447 & .547\sd{.016} & .714\sd{.006} & .492\sd{.007} & .583\sd{.007} \\
\midrule
Best entry & .280\sd{.007} & .335\sd{.007} & .541\sd{.019} & .678\sd{.010} & .392\sd{.044} & .487\sd{.045} \\
Mean, with desc. & .427\sd{.006} & .490\sd{.011} & .512\sd{.008} & .648\sd{.026} & .379\sd{.031} & .525\sd{.022} \\
Mean, exp.\ only & .396\sd{.002} & .455\sd{.005} & .478\sd{.004} & .622\sd{.024} & .367\sd{.031} & .496\sd{.062} \\
\bottomrule
\end{tabular}
\caption{More experience slots cost held-out accuracy without raising labeled accuracy, and mean aggregation shrinks but does not remove the shared index's held-out loss. Jina-v2 and Llama-3.1-8B on the per-seed rebuild of Tab.~\ref{tab:twoindex}, seeds 42/123/456. Top: description + experience slots. Bottom: the shared index, each tool ranked by its best entry (the walk-down rule used elsewhere) or by the mean of its entries, with or without its description entry. Labeled: the 340 labeled queries with no held-out gold tool. Mixed: the 80 multi-tool queries with some gold tools held out.}
\label{tab:budget}
\end{table*}

\paragraph{Slot budget.} After the main runs we swept the two-index router's budget under Jina-v2 on the per-seed rebuild behind Tab.~\ref{tab:twoindex}. On that rebuild, all 2,187 three-plus-two candidate lists and reranker predictions match the saved run, and five description slots reproduce \docs{} exactly. Table~\ref{tab:budget} gives the results. Against descriptions, held-out accuracy changes by $-1.3$, $-1.5$, $-4.4$ and $-10.6$ points with one to four experience slots (query-clustered 95\% intervals $[-3.1,+0.4]$, $[-3.9,+0.9]$, $[-7.3,-1.6]$, $[-14.7,-6.8]$). On the 340 labeled queries with no held-out gold tool the change is $+4.4$, $+4.0$, $+3.5$ and $+3.5$ points ($[+2.1,+6.8]$, $[+1.1,+7.2]$, $[+0.1,+7.0]$, $[-0.6,+7.6]$). One experience slot therefore gives about the labeled gain of two at a slightly smaller held-out cost; both differences lie within the intervals. Three-plus-two scores below four-plus-one on every stratum, so the fixed budget is not a value tuned to this split. Tool-clustered intervals agree, except that the two-plus-three labeled gain reaches zero when clustered by tool ($[0.0,+7.0]$). We swept the budget under Jina-v2 only.

\paragraph{Tool-score aggregation.} The walk-down rule ranks a tool in the shared index by its best entry. A tool with many experience entries thus has many chances to rank high. To test whether this choice produces the held-out loss, we reranked the same per-seed rebuild with each tool scored by the mean cosine of its entries, either over all of them (description included) or over its experience entries only, falling back to the description score for tools without experience (Tab.~\ref{tab:budget}). The reranker sees each selected tool's two highest-scoring entries. On this rebuild the best-entry shared index loses 25.4 held-out points to descriptions. The mean with descriptions included recovers 14.7 of them ($[+11.2,+18.3]$; tool-clustered $[+10.9,+18.5]$) and still loses 10.7 ($[-14.7,-6.9]$). It stays 9.2 points below the two-index router on held-out queries ($[-12.7,-5.7]$). On the 340 labeled queries with no held-out gold tool it ties descriptions ($0.0$, $[-4.8,+4.7]$), so it gives up the labeled gain that experience brings. On mixed queries it loses 19.6 points to descriptions. The mean over experience entries only does worse on every stratum. Aggregation by best entry therefore adds to the held-out loss but does not explain all of it.

\paragraph{The benchmark's own retriever.} The ToolBench-IR row of Tab.~\ref{tab:retrievers} runs the retrieval stage under the retriever released with ToolBench, in the same harness as the off-the-shelf retrievers. That retriever is BERT-base fine-tuned for API retrieval on the ToolBench instruction pool. From descriptions alone it reaches 0.974 of held-out gold tools and 0.967 of labeled ones (0.961 and 0.974 with documents formatted as its training documents were). The shared index gives 0.924\sd{.007} held-out and 0.975\sd{.004} labeled. The crowding signature survives a retriever trained for the task, at a cost of 5 points instead of 26--35. On the independent pool the same retriever reaches 0.005\sd{.001} of held-out gold tools from descriptions (chance 0.001). It still matches labeled queries to pooled scenarios at 0.437\sd{.044}, against Jina's 0.48--0.53. It does not collapse onto a few tools: no top-5 set recurs for more than 1.5\% of held-out queries. On this pool the model fails to match queries to tool descriptions it was not trained on. Whether that reflects memorization or domain shift, we cannot say.

\paragraph{Relative losses.} Relative losses correct for low baseline recall. In these terms, the shared index loses 34--57\% of what descriptions reach under Jina, BGE, and BM25 on the independent pool, and 34--51\% on the split. On the independent pool, only Qwen3 (13\%) and the clean model (1\%) lose much less.

\paragraph{Raw-pool training also hides crowding.} The shared index costs the raw and label-kept models of \S\ref{sec:split} 4.2 and 4.0 points, against 7.1 for the clean one. ToolBench-IR, the benchmark's released retriever, shows a similarly small loss (0.974 to 0.924). A held-out number from such a retriever overstates both its routing to unseen tools and how well that routing withstands pooled experience. ToolBench-IR belongs with the raw-trained models. Its architecture differs, though, so the comparison is not causal for it.

\begin{table}[t]
\centering\footnotesize
\setlength{\tabcolsep}{3pt}
\begin{tabular}{@{}lcccc@{}}
\toprule
 & \multicolumn{2}{c}{cos $\ge 0.95$} & \multicolumn{2}{c}{cos $\ge 0.90$} \\
\cmidrule(lr){2-3}\cmidrule(lr){4-5}
Arm & acc & R@5 & acc & R@5 \\
\midrule
\docs & .528 & -- & .528 & .672 \\
\synapse & .587 & -- & .480 & .649 \\
\central & .585 & -- & .479 & .647 \\
\concat & .585\sd{.007} & .724 & .532\sd{.010} & .687\sd{.005} \\
\clf & .817 & -- & .774\sd{.032} & -- \\
\bottomrule
\end{tabular}
\caption{Overall accuracy with the leak filter at cos $\ge 0.95$ and $\ge 0.90$. At 0.90, \synapse{} and \central{} are seed 42 only.}
\label{tab:threshold}
\end{table}

\paragraph{Retriever fine-tuned on the label-held-out pool (clean model).} We fine-tune BGE-base-en v1.5 with an in-batch-negative contrastive loss on (query, gold-tool description) pairs from the label-held-out pool: 90,704 items, 137,396 expanded pairs (one pair per gold tool of a multi-tool item). Pair and update counts for every training pool are in Tab.~\ref{tab:pairs}. Before training, the run asserts three things: no pair's tool is in $H$, no training query matches a test query exactly, and none has cosine $\ge 0.95$ to a test query under the base model. The last check found 4 queries that the Jina-based leak filter had kept. We removed and logged them. That leaves 137,392 pairs. Training uses one epoch, batch 128, learning rate $2\times10^{-5}$, 10\% warmup, maximum length 256, and three training seeds. On a 2\% development split, recall@5 rose from 0.45 to 0.81 (seed 42). We evaluate the checkpoint in the same harness as Tab.~\ref{tab:retrievers}, with the five-distinct-tools walk-down. From descriptions it gives 0.878\sd{.004} held-out and 0.930\sd{.006} labeled recall@5. The shared index gives 0.807\sd{.002} and 0.924\sd{.009}. On the independent pool (three draws), descriptions give 0.219\sd{.039} held-out and 0.237\sd{.022} labeled recall@5. There the shared index gives 0.216\sd{.038} and 0.298\sd{.030}. For held-out recall, the query-clustered 95\% interval on the difference is $[-0.50,-0.19]$ points.

\begin{table}[t]
\centering\scriptsize
\setlength{\tabcolsep}{2.5pt}
\begin{tabular}{@{}lcccccc@{}}
\toprule
Retriever & gold desc. & nearest exp. & exp.$-$gold & exp.$>$gold & loss \\
\midrule
Jina-v2 & 0.798 & 0.839 & $+$0.041 & .784 & 35.4 \\
Qwen3-Emb-8B & 0.823 & 0.861 & $+$0.038 & .710 & 33.6 \\
BM25$^{\dagger}$ & 43.46 & 51.99 & $+$8.52 & .666 & 26.8 \\
BGE-base & 0.692 & 0.709 & $+$0.018 & .586 & 26.2 \\
BGE, clean & 0.665 & 0.602 & $-$0.063 & .360 & 7.1 \\
ToolRet-e5 & 0.709 & 0.649 & $-$0.060 & .222 & 1.7 \\
\bottomrule
\end{tabular}
\caption{Experience outscores the gold description more often under general-purpose retrievers than under tool-trained ones, in the order of their crowding loss. Similarity gap on the 309 held-out queries, means over three seeds. Gold desc.: query score against its gold description; nearest exp.: against the top pooled-experience entry (cosine; $^{\dagger}$raw BM25 score). exp.$>$gold: share of queries where experience scores higher. Loss: recall@5 points lost under the shared index. Cosine scales differ, so rows are compared by the share.}
\label{tab:gap}
\end{table}

\paragraph{Retriever fine-tuned on the raw pool.} The recipe, seeds, and evaluation harness are those of the clean model. Only the training pool changes, to the post-junk raw pool of 196,023 items. That pool keeps the 745 items that match a test query exactly, 21,377 items labeled with a held-out tool, and 104,875 items that mention one. Held-out recall@5 is 0.962\sd{.005} from descriptions and 0.920\sd{.009} from the shared index. Against the clean model on descriptions (927 query$\times$seed pairs), the raw model alone surfaces 82 held-out queries and the clean model alone surfaces 4 ($p=5.8\times10^{-20}$). The query-clustered 95\% interval on the difference is $[+5.61,+11.33]$ points. On the independent pool (three draws) the raw model gives 0.218\sd{.041} from descriptions and 0.213\sd{.039} from the shared index, against 0.219\sd{.039} and 0.216\sd{.038} for the clean model. On descriptions there are 138/146 discordant pairs ($p=0.678$; interval $[-0.76,+0.50]$ points).

\paragraph{Size-matched raw pool.} The raw pool is about twice the size of the label-held-out pool. We therefore also trained the same recipe on a uniform sample of the raw pool with the label-held-out pool's size (90,704 items, three seeds). The sample has no stratification or forced retention. The sample contains 9,870 held-out-tool-labeled items and 353 exact test-query overlaps. It reaches 0.944\sd{.012} held-out recall@5 from descriptions, $+6.58$ points over the clean model (McNemar 65/4, $p=3.1\times10^{-15}$; interval $[+4.21,+9.17]$). The full raw pool adds a further $+1.83$ (26/9, $p=0.006$; $[+0.32,+3.56]$). At equal size, then, about four fifths of the gain remains. Volume alone cannot account for the remainder. The full pool also holds 2.2 times as many held-out-tool-labeled items (21,377 against 9,870) and 745 exact test-query overlaps against 353. There is no gain on the independent pool either. We therefore attribute the in-pool gain to exposure to the held-out tools' items and to the test queries. This comparison removes the two together.

\paragraph{Label-kept pool.} To separate those two exposures, we trained the same recipe and seeds on a uniform 90,704-item sample of the leak-filtered pool, from which exact and cosine $\ge 0.95$ test-query matches are removed. We drew it without excluding items that label or mention held-out tools. It holds on average 9,482 held-out-labeled items (9,406--9,614 across seeds, against 9,870 in the size-matched raw sample), 48,291 mentioning items, and 0 exact test-query overlaps. Held-out recall@5 from descriptions is 0.930\sd{.005} (shared index 0.890\sd{.009}). Labeled recall is 0.940\sd{.007} (0.947\sd{.008} from the shared index). Against the clean model the gain is $+5.18$ points (McNemar 56/8, $p=5.6\times10^{-10}$; query-clustered interval $[+2.91,+7.55]$, tool-clustered $[+2.82,+7.71]$). The size-matched raw model adds $+1.40$ over it (20/7, $p=0.019$; $[+0.22,+2.70]$ and $[+0.11,+2.88]$). The full raw model adds $+3.24$ (36/6, $p=2.8\times10^{-6}$; $[+1.62,+5.07]$ and $[+1.42,+5.31]$). On the independent pool the label-kept model gives 0.221\sd{.041} from descriptions and 0.216\sd{.040} from the shared index. The size-matched raw model shows no gain there either: 0.218\sd{.036} from descriptions and 0.214\sd{.034} from the shared index, against the clean model's 0.219 (difference $-0.09$ points, $[-0.61,+0.46]$). We release inputs, checkpoints, and evaluation summaries with content hashes.

\paragraph{What the size-matched comparisons control.} The 5.2/1.4 split compares exposure conditions. It is not a causal partition between labels and test queries. The three size-matched pools share their item count but not their pair counts, so at one epoch their optimizer steps differ too (Tab.~\ref{tab:pairs}). The label-kept models took 1,130--1,132 updates and the size-matched raw model took 1,127, against 1,051 for the clean model. The 1.4-point step from label-kept to size-matched raw is therefore matched in updates to within 0.4\%. The 5.2-point step is not (about 7.6\% more updates). The label-kept contrast moves labels and mentions together. The size-matched raw sample differs from the label-kept pool in near-duplicate as well as exact test-query matches. It also holds about 4\% more held-out-labeled items. The label-kept pool is filtered at cosine $0.95$, so test-query paraphrases between 0.90 and 0.95 remain in it. The 0.90 runs below retrain the clean, label-kept, and size-matched raw models at matched updates.

\paragraph{Clean model end to end.} With a tool-trained retriever, the two-index router is the best arm on held-out queries and the worst on labeled ones (Tabs.~\ref{tab:bgee2e} and~\ref{tab:bgearms}). The shared index's loss has intervals that include zero. We run the clean model under descriptions, the shared index, and the two-index router (three description and two experience slots), end to end with the default reranker on the split. On held-out queries the two-index router's 3.56-point gain over descriptions has query- and tool-clustered 95\% intervals of $[+1.19,+6.04]$ and $[+1.36,+5.86]$ (McNemar $p=0.00026$). Its 5.83-point gain over the shared index has $[+3.34,+8.52]$ and $[+2.87,+8.97]$ ($p=1.2\times10^{-8}$). The shared index loses 2.27 points to descriptions, far less than under Jina, with intervals $[-5.29,+0.86]$ and $[-5.48,+0.86]$ ($p=0.055$). Overall accuracy barely moves (0.696 for the two-index router against 0.698 for descriptions). The shared index is 1.0 point below descriptions. The two-index router differs from descriptions by $-0.18$ points and from the shared index by $+0.78$, and both differences have intervals that include zero. The two-index router reaches fewer held-out gold tools than descriptions (recall@5 0.821 against 0.878) but routes more of those it reaches correctly, 0.88 against 0.78 (accuracy over recall@5). These rates condition on different queries and different distractors, so they do not show better reranking. We did not test why the rates differ.

\paragraph{Clean model, labeled queries.} The two-index router's held-out gain costs labeled accuracy. On the 420 labeled queries, accuracy is 0.706\sd{.006} for descriptions, 0.706\sd{.011} for the shared index, and 0.676\sd{.013} for the two-index router. The shared index ties descriptions on average ($+0.00$; query- and tool-clustered intervals $[-2.86,+2.94]$ and $[-3.53,+3.90]$; McNemar discordant pairs 114/114, $p=1.0$). The two-index router loses 2.94 points to descriptions ($[-5.24,-0.79]$ and $[-5.08,-0.72]$; 88/51, $p=0.0022$). It also loses 2.94 points to the shared index ($[-5.48,-0.56]$ and $[-6.42,+0.34]$; 108/71, $p=0.0070$). Labeled recall@5 is 0.930 for both descriptions and the two-index router, so the loss is not a recall@5 loss. We did not test where it arises.

\begin{table}[t]
\centering\scriptsize
\setlength{\tabcolsep}{3pt}
\begin{tabular}{@{}lcccc@{}}
\toprule
 & \multicolumn{2}{c}{no held-out gold (340)} & \multicolumn{2}{c}{mixed (80)} \\
\cmidrule(lr){2-3}\cmidrule(lr){4-5}
Arm & acc & R@5 & acc & R@5 \\
\midrule
\docs & .707\sd{.007} & .937\sd{.006} & .700\sd{.025} & .900\sd{.022} \\
Shared index & .724\sd{.013} & .940\sd{.007} & .629\sd{.007} & .854\sd{.031} \\
Two-index router & .679\sd{.025} & .948\sd{.007} & .663\sd{.038} & .854\sd{.029} \\
\bottomrule
\end{tabular}
\caption{The two-index router's labeled loss appears on queries with no held-out gold tool too. The shared index gains there and loses on mixed queries. Clean BGE model end to end on the two parts of the labeled stratum, three seeds. Mixed: multi-tool queries with some gold tools held out.}
\label{tab:strata}
\end{table}

\paragraph{Labeled queries without held-out gold tools.} The 420 labeled queries include 80 mixed multi-tool queries, some of whose gold tools are held out. Table~\ref{tab:strata} separates the two parts in the same runs. We report seed means and ran no paired tests on these parts. The two-index router's labeled loss is not confined to the mixed queries. On the 340 with no held-out gold tool it scores 0.679 against 0.707 for descriptions, although its recall@5 there is higher (0.948 against 0.937). The shared index's tie on all 420 queries combines two opposite parts. It scores 1.7 points above descriptions on the 340. On the 80 mixed queries it scores 7.1 points below them, and its recall@5 also falls (0.854 against 0.900). The mixed-query loss fits crowding of their held-out gold tools, but we did not test that directly.

\paragraph{Clean model at retrieval.} On the split, the two-index router recovers 1.4 of the 7.1 recall points the shared index costs and keeps labeled recall at the description level. On the independent pool it lowers held-out recall by 4.5 points, from 0.219 to 0.174. The shared index costs 0.3 there. Both experience-carrying indices raise labeled recall by about 6 points. We did not route end to end on that pool. The retrieval-only cells report seed means without paired tests.

\paragraph{Clean model at the stricter filter.} To check whether paraphrases between cosine 0.90 and 0.95 drive the clean baseline, we rebuilt the label-held-out pool from the 0.90-filtered pool (87,599 items). We retrained with the same recipe and fixed the update count at the clean run's 1,051. Held-out recall from descriptions is 0.873\sd{.013} against 0.878\sd{.004}. The shared index costs 6.8 points against 7.1. Held-out accuracy keeps the same ordering (0.709 for the two-index router, 0.676 for descriptions, 0.654 for the shared index). The clean baseline does not depend on those paraphrases.

\paragraph{Label-kept pool at the stricter filter.} To test whether residual paraphrases drive the label-kept gain, we sampled 87,599 items from the 0.90-filtered pool without removing held-out-tool items. That is the size of the 0.90 clean pool. The sample has 0 exact test-query overlaps and 7,961, 8,083, and 7,934 held-out-labeled items per seed. Both models train for exactly 1,051 updates. The base-model leak check dropped 41, 38, and 36 queries (48, 41, and 42 pairs) per seed. Held-out recall@5 from descriptions is 0.929\sd{.012} against the 0.90 clean model's 0.873\sd{.013}. The 5.6-point gap is close to the 5.2 points at the 0.95 filter (Tab.~\ref{tab:bgearms}). On the independent pool the label-kept model gives 0.224\sd{.040} against 0.219\sd{.037}, so the gain stays specific to the held-out tools. Paired against the 0.90 clean model, the label-kept model leads under all three indices. It leads by 5.61 points from descriptions (query- and tool-clustered 95\% intervals $[+3.34,+8.09]$ and $[+3.19,+8.20]$; McNemar $p=2.4\times10^{-11}$). From the shared index it leads by 7.44 points ($[+4.64,+10.36]$, $[+4.44,+10.60]$; $p=9.8\times10^{-14}$). Under the two-index router it leads by 6.69 ($[+4.10,+9.39]$, $[+3.95,+9.65]$; $p=1.7\times10^{-13}$). The run still moves labels and mentions together.

\paragraph{Size-matched raw pool at the stricter filter.} To rerun the 1.4-point step at 0.90, we took the size-matched raw sample (90,704 items). We removed the 944 items that match a test query at cosine $\ge 0.90$ without matching it exactly. The 353 exact test-query items (514 pairs) stay in. That leaves 89,760 items and 146,235 pairs, which we trained on for 1,051 updates with the same recipe and seeds. A first version also dropped the exact matches in the base-model leak check. That version no longer had the exposure under test, so we stopped it and report only the corrected run. Under the split's mention rule the pool holds 9,075 held-out-labeled and 52,543 mentioning items, against 7,934--8,083 and 45,945--46,019 in the label-kept 0.90 pools. It therefore differs from them in three ways at once: exact test-query items, about 14\% more held-out labels, and about 14\% more mentions. Held-out recall@5 from descriptions is 0.934\sd{.005} (shared index 0.888\sd{.004}, two-index router 0.878\sd{.008}). On the independent pool it is 0.221\sd{.043}. Against the label-kept 0.90 model the difference is $+0.54$ points (McNemar 15/10, $p=0.42$; query- and tool-clustered 95\% intervals $[-0.54,+1.62]$ and $[-0.66,+2.04]$). Against the clean 0.90 model it is $+6.15$ (63/6, $p=4.5\times10^{-13}$; $[+3.88,+8.74]$ and $[+3.49,+9.17]$). At 0.90, then, the exact test queries, together with the extra held-out labels and mentions, add no detectable recall over the label-kept pool. The size-matched raw sample at 0.95 had no filter at all, so it also kept test-query paraphrases. We did not test whether those produce the 1.4-point step there.

\begin{table}[t]
\centering\scriptsize
\setlength{\tabcolsep}{2.5pt}
\begin{tabular}{@{}llcccc@{}}
\toprule
Filter & Arm & \multicolumn{3}{c}{accuracy} & held-out \\
 & & held-out & labeled & all & R@5 \\
\midrule
0.95 & \docs & .687\sd{.010} & .706\sd{.006} & .698\sd{.006} & .878 \\
0.95 & shared index & .665\sd{.019} & .706\sd{.011} & .688\sd{.013} & .807 \\
0.95 & two-index router & .723\sd{.005} & .676\sd{.013} & .696\sd{.007} & .821 \\
\midrule
0.90 & \docs & .676\sd{.015} & -- & .692\sd{.013} & .873 \\
0.90 & shared index & .654\sd{.010} & -- & .682\sd{.002} & .805 \\
0.90 & two-index router & .709\sd{.006} & -- & .690\sd{.009} & .808 \\
\bottomrule
\end{tabular}
\caption{At both filters the two-index router leads on held-out accuracy, and overall accuracy stays within one point across arms. Clean BGE model end to end on the split, three training seeds. Filter: leak-filter threshold of the training pool. Labeled accuracy is not reported separately for the 0.90 rows; R@5 from Tab.~\ref{tab:bgearms}.}
\label{tab:bgee2e}
\end{table}

\begin{table*}[t]
\centering\small
\setlength{\tabcolsep}{2.6pt}
\begin{tabular}{@{}llcccccc@{}}
\toprule
 & & \multicolumn{2}{c}{descriptions} & \multicolumn{2}{c}{shared index} & \multicolumn{2}{c}{two-index router} \\
\cmidrule(lr){3-4}\cmidrule(lr){5-6}\cmidrule(lr){7-8}
Training filter & Pool & held-out & labeled & held-out & labeled & held-out & labeled \\
\midrule
cos $\ge 0.95$ & Split & .878\sd{.004} & .930\sd{.005} & .807\sd{.002} & .924\sd{.009} & .821\sd{.002} & .930\sd{.005} \\
cos $\ge 0.95$ & Indep. & .219\sd{.039} & .237\sd{.022} & .216\sd{.038} & .298\sd{.030} & .174\sd{.039} & .296\sd{.028} \\
cos $\ge 0.90$ & Split & .873\sd{.013} & .917\sd{.004} & .805\sd{.011} & .924\sd{.010} & .808\sd{.005} & .927\sd{.010} \\
cos $\ge 0.90$ & Indep. & .219\sd{.037} & .240\sd{.023} & .215\sd{.035} & .299\sd{.032} & .180\sd{.035} & .296\sd{.031} \\
\midrule
Label-kept, 0.90 & Split & .929\sd{.012} & -- & .879\sd{.013} & -- & .875\sd{.008} & -- \\
Label-kept, 0.90 & Indep. & .224\sd{.040} & -- & .217\sd{.038} & -- & .181\sd{.037} & -- \\
Raw size-m., 0.90 & Split & .934\sd{.005} & .939\sd{.007} & .888\sd{.004} & .954\sd{.018} & .878\sd{.008} & .956\sd{.013} \\
Raw size-m., 0.90 & Indep. & .221\sd{.043} & -- & .215\sd{.042} & -- & .177\sd{.043} & -- \\
\bottomrule
\end{tabular}
\caption{Pooled experience costs the clean model about 7 held-out recall points on the split and almost none on the independent pool. Label-kept and raw training raise split recall but not independent-pool recall. Clean BGE model, recall@5 under three candidate indices, retrieval only; three training seeds (independent pool: three draws). 0.95 rows: the clean model of Tab.~\ref{tab:retrievers}. 0.90 rows: models retrained at 1,051 updates on pools rebuilt from the 0.90-filtered pool (App.~\ref{app:setup}).}
\label{tab:bgearms}
\end{table*}

\begin{table*}[t]
\centering\small
\setlength{\tabcolsep}{6pt}
\begin{tabular}{@{}lrrrr@{}}
\toprule
Pool & items & pairs & train & updates \\
\midrule
Clean (0.95) & 90,704 & 137,392 & 134,645 & 1,051 \\
Label-kept & 90,704 & 147,640--147,924 & 144,688--144,966 & 1,130--1,132 \\
Size-m.\ raw & 90,704 & 147,302 & 144,356 & 1,127 \\
Full raw & 196,023 & 318,412 & 312,044 & 2,437 \\
Clean (0.90) & 87,599 & 133,068 & 130,407 & 1,051$^{f}$ \\
Label-kept (0.90) & 87,599 & 143,430--143,658 & 140,562--140,785 & 1,051$^{f}$ \\
Size-m.\ raw (0.90) & 89,760 & 146,235 & 143,311 & 1,051$^{f}$ \\
\bottomrule
\end{tabular}
\caption{Training-pool sizes for the BGE recipe; ranges over seeds. Pairs: after multi-tool expansion and the base-model leak check. Train: after the 2\% development split. Updates: at batch 128 for one epoch, or fixed ($^{f}$) to the clean run's.}
\label{tab:pairs}
\end{table*}

\begin{table}[t]
\centering\footnotesize
\setlength{\tabcolsep}{3pt}
\begin{tabular}{@{}lcccc@{}}
\toprule
 & \multicolumn{2}{c}{held-out (310)} & \multicolumn{2}{c}{labeled (358)} \\
\cmidrule(lr){2-3}\cmidrule(lr){4-5}
Retriever & docs & shared & docs & shared \\
\midrule
Jina-v2 & .639 & .352\sd{.009} & .707 & .676\sd{.007} \\
BGE-base & .732 & .539\sd{.006} & .788 & .755\sd{.004} \\
BM25 & .677 & .566\sd{.007} & .715 & .809\sd{.006} \\
\midrule
\multicolumn{5}{@{}l}{Jina routing accuracy} \\
\docs & .506 & & .550 & \\
\concat & & .289 & & .581 \\
Two-index router & & .477\sd{.006} & & .594\sd{.006} \\
\bottomrule
\end{tabular}
\caption{Holding out whole categories also produces a held-out recall loss under all three retrievers, smaller than on the split. Category-level hold-out (738 tools; pool 42,648 items): recall@5 and Jina routing accuracy, seeds 42/123/456. Shared: descriptions plus pooled experience, concatenated (\concat).}
\label{tab:e7}
\end{table}

\paragraph{Category-level hold-out.} The split holds out 172 tools but leaves their category neighbours labeled. Here $\mathcal{C}$ is the set of the seven categories that contain a G1-category gold tool. We hold out every catalog tool in $\mathcal{C}$ (738). We remove a pool item if any label is in $\mathcal{C}$, if its attached category is in $\mathcal{C}$, or under the split's mention rule. That leaves 42,648 items, above the 25,000 needed for five 5,000-item client draws. Held-out queries are those whose gold tools are all held out (310). Labeled queries have none held out (358), and we report the 61 mixed queries separately. We use \concat in place of \synapse as the pooled-experience arm. We recorded this substitution before the run, on the strength of their equivalence at the 0.95 filter (App.~\ref{app:controls}), and it avoids the edge merge. In Table~\ref{tab:e7}, the shared index loses held-out recall under all three retrievers (28.7, 19.3, and 11.1 points under Jina, BGE, BM25, against 35.4, 26.2, and 26.8 on the split). Labeled recall falls under Jina and BGE and rises under BM25. The pool is less than half the split's, and the client partition differs. We therefore compare arms within this construction only. Held-out McNemar tests against \docs{} (930 pairs) give 267/0 for \concat{} ($p=8.4\times10^{-81}$) and 69/0 for the two-index router ($p=3.4\times10^{-21}$). On labeled queries the two-index router against \docs{} gives 39/114 ($p=1.0\times10^{-9}$).

\begin{table}[t]
\centering\scriptsize
\setlength{\tabcolsep}{2pt}
\begin{tabular}{@{}lcccccc@{}}
\toprule
 & \multicolumn{2}{c}{one owner ($T_x$)} & \multicolumn{2}{c}{held-out ($H$)} & \multicolumn{2}{c}{other labeled} \\
\cmidrule(lr){2-3}\cmidrule(lr){4-5}\cmidrule(lr){6-7}
Arm & acc & R@5 & acc & R@5 & acc & R@5 \\
\midrule
\docs & .544 & .684 & .534 & .689 & .529 & .673 \\
\local{} & .425\sd{.005} & .520\sd{.003} & .411 & .494 & .354 & .438 \\
\synapse & .605\sd{.028} & .745\sd{.024} & .293 & .345 & .496 & .615 \\
\concat & .600\sd{.011} & .745\sd{.019} & .294 & .345 & .498 & .611 \\
Two-index router & .566\sd{.013} & .750\sd{.007} & .505 & .612 & .572 & .738 \\
\bottomrule
\end{tabular}
\caption{On tools one other client has used, pooled experience beats each client's own compendium. Cross-client construction, Jina routing, seeds 42/123/456. $T_x$: the tools whose labeled pool items are held by exactly one client (the owner). Scores are on the 136 queries whose gold tool is in $T_x$, evaluated at the four non-owners. \local{} is the evaluating client's own compendium.}
\label{tab:e8}
\end{table}

\paragraph{Cross-client construction.} We start from the split's pool, so $H$ stays unlabeled everywhere. The gold tools of the G1-instruction queries that are not in $H$ form $T_x$. For each seed, a seeded hash assigns each $t\in T_x$ an owner client. Every pool item labeled $t$ goes into the owner's 5,000-item draw and is removed from the other clients' draws. We refill their vacated slots with items labeled with no tool in $T_x$. We drop any item whose tools have different owners. Of 144 candidate tools, 129 had items that could be assigned to a single owner and form $T_x$. The other 15 had none and are excluded. We evaluate the 136 queries whose gold tool is in $T_x$ at each of the four non-owner clients. Means are therefore over (query, client) pairs, and the declared tests cluster by query. The run asserts that exactly one client holds items labeled $t$ for every $t\in T_x$. It also asserts that \docs{} reproduces its earlier per-query top-5 lists, predictions, and recall on all 729 queries for every seed. Owners hold a median of 8 labeled items per tool (IQR 8--8, range 1--8). Non-owners still see some text about these tools. Per seed, between 1,822 and 1,904 of their items mention a $T_x$ tool by name but are labeled with another. For the pooled arms and \docs{}, the prediction for a query does not depend on the evaluating client. The four non-owner rows of a query are then identical, and McNemar tests over rows would overstate significance. As declared before the run, we read this experiment from the query-clustered intervals. $T_x$ means in Tab.~\ref{tab:e8} are over the 1,632 (query, client, seed) rows of the declared test population. By seed, \synapse{} scores 0.618, 0.574, and 0.625. Descriptions score 0.544. The +6.1-point gain of pooled experience over them has the interval $[-0.5,+12.7]$. The two-index router falls 3.9 points short of \synapse ($[-9.3,+1.5]$). With an interval half-width of about 6.6 points, this test cannot see gains over descriptions below about 7 points.

\paragraph{Near-duplicate threshold.} We rebuilt the pool with the filter at cosine $\ge 0.90$. The filter removes 745 exact and 11,055 near-duplicate items from the raw pool of 196,023, which leaves 184,223. The run asserts zero exact and zero cosine $\ge 0.90$ matches between test queries and pool. Clients draw 5,000 items from this 0.90-filtered pool with the same seed and configuration. In Table~\ref{tab:threshold}, \docs reproduces its three-seed value exactly. \synapse and \central (seed 42) lose about 11 points and fall 4.8 points below descriptions. \concat needs no edge merge, so we ran it on all three seeds. It loses 5.3 points and ends level with descriptions (0.532 against 0.528; 0.538, 0.520, and 0.538 on seeds 42, 123, and 456). On seed 42, where both were run, \synapse and \concat differ by 5.8 points (0.480 against 0.538). Their equivalence is established at the 0.95 filter only, and we do not test it here. The experience arms' lead over descriptions does not survive the tighter filter. Whether they fall below descriptions depends on the arm. The \concat runs use a walk-down cap of 5,000, because one seed-123 query could not be filled with five distinct tools within 200 entries. \concat keeps every client's copy of a scenario, so one tool can occupy many entries. We refit the classifier on each seed's five client draws. The run asserts that the fit sets have no exact or cosine $\ge 0.90$ matches to the test queries and that the pool hash is identical across seeds. The classifier scores 0.755, 0.811, and 0.757 on seeds 42, 123, and 456 (0.774\sd{.032}). It loses about 4 points to the tighter filter, so part of its margin at 0.95 was paraphrase too. Its seed-to-seed variation at this threshold is larger than the roughly one-point SD of the LLM arms elsewhere. That is one more reason to read the single-seed experience-arm cells only for the size of their drop.

\begin{table*}[t]
\centering\small
\setlength{\tabcolsep}{3pt}
\begin{tabular}{@{}llcccccc@{}}
\toprule
 & & \multicolumn{3}{c}{held-out R@5} & \multicolumn{3}{c}{labeled R@5} \\
\cmidrule(lr){3-5}\cmidrule(lr){6-8}
Expansion & Retriever & synth. & +exp. & capped & synth. & +exp. & capped \\
\midrule
Template & Jina-v2 & .706 & .357\sd{.005} & .387\sd{.005} & .667 & .653\sd{.010} & .613\sd{.028} \\
Template & BGE-base & .728 & .553\sd{.006} & .565\sd{.008} & .733 & .737\sd{.006} & .720\sd{.008} \\
Template & BM25 & .699 & .358\sd{.007} & .376\sd{.005} & .757 & .784\sd{.010} & .763\sd{.014} \\
\midrule
Generated & Jina-v2 & .602 & .462\sd{.002} & .547\sd{.009} & .543 & .685\sd{.007} & .633\sd{.010} \\
Generated & BGE-base & .615 & .608\sd{.003} & .696\sd{.006} & .600 & .670\sd{.006} & .655\sd{.013} \\
Generated & BM25 & .660 & .488\sd{.008} & .530\sd{.008} & .624 & .819\sd{.018} & .802\sd{.007} \\
\midrule
Gen., indep. & Jina-v2 & .107\sd{.022} & .102\sd{.024} & .103\sd{.024} & -- & -- & -- \\
Gen., indep. & BGE-base & .092\sd{.031} & .087\sd{.030} & .087\sd{.030} & -- & -- & -- \\
Gen., indep. & BM25 & .118\sd{.004} & .102\sd{.003} & .103\sd{.005} & -- & -- & -- \\
\midrule
\multicolumn{2}{@{}l}{No expansion (Tab.~\ref{tab:retrievers})} & \multicolumn{6}{l}{descriptions .689 / .764 / .689; shared index .335 / .502 / .421 (Jina / BGE / BM25)} \\
\bottomrule
\end{tabular}
\caption{Under neither template nor generated expansion does an index with pooled experience reach the held-out recall of the description index. Every catalog tool is indexed with five queries written from its description alone, by template or by a frozen Llama-3.1-8B-Instruct (Generated). Synth.: descriptions plus expansion. +exp.: plus pooled experience. Capped: five non-description entries per tool. Gen., indep.: independent pool (three draws); other rows on the split.}
\label{tab:symmetric}
\end{table*}

\paragraph{Symmetric expansion.} Template expansion alone is roughly neutral for held-out tools (Tab.~\ref{tab:symmetric}). Adding pooled experience on top collapses held-out recall to within a few points of the unexpanded shared index. Capping every tool at the same entry count does not change that. Generated expansion lowers held-out recall on its own under all three retrievers (by 8.7, 14.9, and 2.9 points against descriptions under Jina, BGE, BM25). It lowers labeled recall as well. These runs use the same five-distinct-tools walk-down (pool 200; no query had fewer than five distinct tools), so the drop is not underfilling. Adding pooled experience costs 14.0, 0.65, and 17.3 points of held-out recall against 34.9, 17.5, and 34.1 under template expansion. Under BGE, capping raises held-out recall above the generated-only index (0.696 against 0.615). Under Jina and BM25 it does not. All three indices hold the 3,180 canonical descriptions. The synthetic index adds the 15,900 generated queries. The +exp.\ index adds every pooled experience entry on top. The capped index adds exactly five non-description entries per tool. It takes real experience first and fills the remaining slots with generated queries. Per seed it keeps 10,495, 10,609, and 10,424 experience entries and 5,405, 5,291, and 5,476 generated ones. It drops 23,358, 24,027, and 23,788 experience entries above the cap. In a dense index, entries added for other tools can only push a held-out tool down. The capped BGE gain therefore comes from replacing generated entries with real experience and truncating surplus experience, not from adding experience. The generated queries have 0 exact and 0 cosine $\ge 0.95$ matches to the 729 test queries, and 41 of the 15,900 lie at cosine $\ge 0.90$. The synthetic-only columns are deterministic. On the independent pool we generated 25,700 queries the same way, with 0 exact and 0 cosine $\ge 0.90$ matches to its evaluation queries.

\begin{table*}[t]
\centering\small
\setlength{\tabcolsep}{2.6pt}
\begin{tabular}{@{}lcccccc@{}}
\toprule
 & \multicolumn{2}{c}{all 729} & \multicolumn{2}{c}{held-out} & \multicolumn{2}{c}{labeled} \\
\cmidrule(lr){2-3}\cmidrule(lr){4-5}\cmidrule(lr){6-7}
Candidate index & acc & R@5 & acc & R@5 & acc & R@5 \\
\midrule
Descriptions & .667 & .967 & .650 & .961 & .679 & .971 \\
Shared & .523\sd{.009} & .725\sd{.007} & .400\sd{.010} & .558\sd{.008} & .613\sd{.009} & .848\sd{.006} \\
\bottomrule
\end{tabular}
\caption{Even at 200 distinct tools the shared index reaches far fewer held-out gold tools than descriptions. Bounded-depth gold insertion, three seeds. Each index is walked to 200 distinct tools, and the gold tool replaces the lowest-ranked non-gold candidate only if reachable there. Acc: the reranker after insertion; R@5: the perfect-selection ceiling. Not comparable cell by cell with Tab.~\ref{tab:loss}.}
\label{tab:bounded}
\end{table*}

\begin{table}[t]
\centering\scriptsize
\setlength{\tabcolsep}{3pt}
\begin{tabular}{@{}lcccc@{}}
\toprule
 & \multicolumn{2}{c}{held-out} & \multicolumn{2}{c}{labeled} \\
\cmidrule(lr){2-3}\cmidrule(lr){4-5}
Candidates from & docs & comp. & docs & comp. \\
\midrule
description index & .540 & .534 & .531 & .524 \\
shared index & .296 & .289 & .524 & .517 \\
\bottomrule
\end{tabular}
\caption{Which index proposes the candidates decides held-out accuracy. What the reranker reads about them barely matters. Rows fix the candidate source and columns the rendering (descriptions only, or the compendium). Real retrieval, mean over three seeds; the difference-in-differences on all 729 queries is 0.000.}
\label{tab:cross}
\end{table}

\paragraph{Candidate source and rendering.} Table~\ref{tab:cross} varies the two factors independently on the same queries and seeds. Rendering costs 0.7 of a point from either index. The index costs 24 points on held-out queries under either rendering, and the difference-in-differences is 0.000. On labeled queries the two indices are level (0.531 against 0.524). When retrieval does surface a gold tool, the shared index turns it into a correct route more often than the description index does (0.83 against 0.78 in Tab.~\ref{tab:heldout}). Because the two rates condition on different queries, this establishes neither a reranking penalty nor better reranking.

\paragraph{Bounded-depth gold insertion.} In Table~\ref{tab:bounded}, descriptions reach the gold tool within 200 tools for 0.967 of all queries and 0.961 of held-out ones. The shared index reaches it for 0.725\sd{.007} and 0.558\sd{.008}. The insertion is therefore available on 0.443 of queries for descriptions but only 0.294\sd{.019} for the shared index. For the rest, the shared index does not hold the gold tool within two hundred tools. Reach bounds a perfect selector over what each index can offer: 0.558 against 0.961 on held-out queries. The current reranker after insertion ends 25 points below descriptions (0.400\sd{.010} against 0.650). Swapping the renderings moves each arm by at most 2.2 points (0.628 and 0.416\sd{.018}). Of the two insertion interventions, this one is closer to what a deployed router could do.

\begin{table}[t]
\centering\scriptsize
\setlength{\tabcolsep}{3pt}
\begin{tabular}{@{}lcc@{}}
\toprule
Difference (held-out) & pts & tool-clustered 95\% CI \\
\midrule
\synapse${}-{}$\docs, accuracy & $-25.6$ & $[-30.8,-20.6]$ \\
Jina, shared${}-{}$docs, R@5 & $-35.4$ & $[-40.9,-29.9]$ \\
Qwen3, shared${}-{}$docs, R@5 & $-33.5$ & $[-39.5,-27.7]$ \\
ToolRet-e5, shared${}-{}$docs, R@5 & $-1.7$ & $[-3.1,-0.6]$ \\
BGE raw${}-{}$clean, R@5 & $+8.4$ & $[+5.1,+12.0]$ \\
BGE size-matched${}-{}$clean, R@5 & $+6.6$ & $[+4.0,+9.4]$ \\
BGE label-kept${}-{}$clean, R@5 & $+5.2$ & $[+2.8,+7.7]$ \\
BGE size-matched${}-{}$label-kept, R@5 & $+1.4$ & $[+0.1,+2.9]$ \\
BGE label-kept${}-{}$clean, 0.90, R@5 & $+5.6$ & $[+3.2,+8.2]$ \\
BGE size-m.\ raw${}-{}$label-kept, 0.90 & $+0.5$ & $[-0.7,+2.0]$ \\
\bottomrule
\end{tabular}
\caption{Tool-clustered intervals for the headline comparisons. Paired bootstrap that resamples the 172 held-out tools (each tool's queries and seed rows kept together; 10,000 resamples).}
\label{tab:toolclust}
\end{table}

\paragraph{Paired tests.} We test differences between arms per query, on the queries both arms share. McNemar's exact two-sided test runs on the paired correct/incorrect (or hit/miss) table. A paired bootstrap 95\% interval on the difference resamples queries with each query's seed rows kept together (10,000 resamples). Equivalence uses two one-sided tests at a 2-point margin declared before any test was run. For \synapse against \concat on all 729 queries, the difference is $+0.14$ points with interval $[-0.55,+0.82]$, inside the margin. Its half-width is 0.69, and the unclustered bootstrap on the same rows gives 0.71. On the 258-query subset the interval is inside the margin at 0\% contradiction and crosses it at 60\%. For the rendering comparisons and the conflict policies, the clustered half-widths reach 2.6 and 2.2 points, and we claim no equivalence. For the headline comparisons we also resample tools in place of queries (172 tool clusters, 10,000 resamples; Tab.~\ref{tab:toolclust}). A multi-tool query is assigned to the cluster of its first-listed gold tool and is not duplicated across its other gold tools. Among held-out queries this affects the 23 multi-tool ones. Tool-clustered intervals lead to the same conclusion as their query-clustered counterparts, with one exception: with the clean model, the two-index router's 2.94-point labeled loss against the shared index excludes zero when clustered by query and includes it when clustered by tool. In Tab.~\ref{tab:retrievers}, every off-the-shelf shared-index cell on the two ToolBench-derived pools lies more than twenty seed SDs below its description cell. On the independent pool the gaps are 5--8 points (two to six draw SDs) for Jina, BGE, and BM25, and 1.5 points for Qwen3-Embedding-8B. The ToolBench-IR cells there are near chance (.005 and .000). Table cells are unweighted means over seeds or draws. McNemar tests pool all held-out query pairs across them (for the independent pool, 5,653 pairs over three draws), so a test's pooled means can differ from the cell in the third decimal. For a held-out tool, a dense shared index can only displace the gold tool and never add it. Under the dense retrievers every discordant pair therefore runs the same way, and $p$ reflects only how many gold tools were displaced. Under BM25, adding documents changes term statistics. A few pairs therefore run the other way (4 of 257 on the split, 402 of 13,934 on the ToolBench single-tool pool, 8 of 452 on the independent pool). We report the counts (Tab.~\ref{tab:mcnemar}), since a very small drop over many pairs can be significant. McNemar tests pool a query's seed rows as separate pairs. Where one arm does not depend on the seed (\docs), this overstates significance. We therefore treat pooled McNemar $p$-values as descriptive and base inference on the query- and tool-clustered intervals. Other tests quoted in the text: two-index router against \docs on held-out accuracy, $p=0.11$; \synapse against \central on all 729 queries, $p=0.79$.

\begin{table}[t]
\centering\scriptsize
\setlength{\tabcolsep}{2pt}
\begin{tabular}{@{}llrrrrl@{}}
\toprule
Pool & Retriever & pairs & docs & shared & d/s & $p$ \\
\midrule
Split & Jina-v2 & 927 & .689 & .335 & 328/0 & $3.7\times10^{-99}$ \\
Split & BGE-base & 927 & .764 & .502 & 243/0 & $1.4\times10^{-73}$ \\
Split & BM25 & 927 & .689 & .421 & 253/4 & $1.6\times10^{-69}$ \\
Split & Qwen3-Emb-8B & 927 & .790 & .454 & 311/0 & $4.8\times10^{-94}$ \\
Split & ToolRet-e5 & 927 & .845 & .827 & 16/0 & $3.1\times10^{-5}$ \\
Split & BGE, clean & 927 & .878 & .807 & 66/0 & $2.7\times10^{-20}$ \\
G1 & Jina-v2 & 71,777 & .763 & .263 & 35,916/0 & $<10^{-300}$ \\
G1 & BGE-base & 71,777 & .847 & .360 & 34,944/0 & $<10^{-300}$ \\
G1 & BM25 & 71,777 & .713 & .530 & 13,532/402 & $<10^{-300}$ \\
Indep. & Jina-v2 & 5,653 & .168 & .112 & 318/0 & $3.7\times10^{-96}$ \\
Indep. & BGE-base & 5,653 & .180 & .111 & 387/0 & $6.3\times10^{-117}$ \\
Indep. & BM25 & 5,653 & .136 & .059 & 444/8 & $7.1\times10^{-120}$ \\
Indep. & Qwen3-Emb-8B & 5,653 & .117 & .102 & 85/0 & $5.2\times10^{-26}$ \\
Indep. & BGE, clean & 5,653 & .217 & .214 & 19/0 & $3.8\times10^{-6}$ \\
Indep. & ToolBench-IR & 5,653 & .005 & .000 & 29/0 & $3.7\times10^{-9}$ \\
\bottomrule
\end{tabular}
\caption{Paired McNemar tests for Tab.~\ref{tab:retrievers} on held-out queries, query$\times$seed (or draw) pairs pooled; descriptive only (App.~\ref{app:setup}). d/s: pairs where only the description index, or only the shared index, surfaced a gold tool.}
\label{tab:mcnemar}
\end{table}

\paragraph{Cost.} The edge merge dominates wall-clock: clustering tens of thousands of scenarios takes roughly an hour per round on an unloaded machine in our implementation. Per-query inference at 3,180 tools is 1.1~ms retrieval and 36.6~ms rerank on one A100-class card. A client's uploaded package is 18,216~KB\,$\pm$\,759 (15 packages, three rounds). Each clean BGE fine-tuning run took about 9 minutes of recorded elapsed time, including evaluation and checkpoint writing; we did not record total GPU-hours across all runs. The released environment pins sentence-transformers 5.6.1 and transformers 5.16.1; BM25 is our own implementation.

\section{The typed merge}
\label{app:merge}

Algorithm~\ref{alg:merge} gives the full edge step. In round $r$, an edge aggregator combines the $K$ client compendiums into one edge compendium $C_E^{(r)}$, written $C_E$ inside the algorithm. Each entry the algorithm emits is one merged unit of $C_E$. A client $k\in\{1,\dots,K\}$ holds a compendium $C_k=(M,U,P,T,A)$ with five parts: tool metadata with usage statistics ($M$), usage scenarios ($U$), precautions ($P$), the API names that ToolBench documents for each tool ($T$), and an annex ($A$). The runner does not synthesize argument templates for $T$. The annex records each item's source group and primary category, or, for a description, its registry source and categories. We write $C.X$ for part $X$ of compendium $C$. A unit $u\in C_k.U$ is one contribution about one tool, $\mathrm{tool}(u)$, and $u.f$ is its value in field $f\in\{$scenario, precaution, annex$\}$. Validation and clustering need three more objects. The schema $S$ gives every field a type and a validation rule. The registry $\mathcal{R}$ lists the catalog's tools with their canonical metadata. Units are clustered at cosine $\ge\theta$, with $\theta=0.85$.
{\small
\begin{algorithm}[htbp]
\caption{\textsc{TypedMerge}: edge merge with field-wise conflict resolution. Policy $\pi\in\{$\textsc{ConflictLog}, \textsc{RoundDelayed}, \textsc{Majority}$\}$. \textsc{ConflictLog} is the \synapse default. Notation is defined in App.~\ref{app:merge}.}
\label{alg:merge}
\begin{algorithmic}[1]
\Require client compendiums $\{C_k\}_{k=1}^K$, schema $S$, registry $\mathcal{R}$, threshold $\theta$, dissent $L^{(r-1)}$ held from round $r-1$ (\textsc{RoundDelayed} only)
\Ensure edge compendium $C_E^{(r)}$
\State reject any $C_k$ that fails validation under $S$; $L^{(r)}\gets\emptyset$
\State $C_E.M \gets$ canonical metadata from $\mathcal{R}$; each numeric subfield $\gets$ mean of the clipped per-client values
\For{each tool $t\in\mathcal{R}$}
  \State $\mathcal{Q}_t \gets$ clusters, at cosine $\ge\theta$, of the units $\{u\in C_k.U : \mathrm{tool}(u)=t,\ k=1,\dots,K\}$
  \For{each cluster $Q\in\mathcal{Q}_t$}
    \State $v_f\gets\textsc{Majority}(\{u.f : u\in Q\})$ for each field $f$
    \State $D_f\gets\{u.f : u\in Q,\ u.f\neq v_f\}$ for each field $f$
    \State $\mathbf{v}\gets(v_{\text{scenario}},v_{\text{precaution}},v_{\text{annex}})$
    \If{$\pi=\textsc{ConflictLog}$} emit $(t,\mathbf{v})$ with conflict log $\{(f,D_f)\}_f$
    \ElsIf{$\pi=\textsc{RoundDelayed}$} emit $(t,\mathbf{v})$; $L^{(r)}\gets L^{(r)}\cup\{(t,f,D_f)\}_f$
    \Else{} emit $(t,\mathbf{v})$ and discard $D_f$
    \EndIf
  \EndFor
\EndFor
\If{$\pi=\textsc{RoundDelayed}$} prepend the entries of $L^{(r-1)}$ to the matching tools' precautions in $C_E.P$
\EndIf
\State $C_E.T\gets$ each tool's API names from $\{C_k.T\}_k$, deduplicated; $C_E.A\gets$ union of $\{C_k.A\}_k$ with duplicate entries removed
\State \Return $C_E^{(r)}$
\end{algorithmic}
\end{algorithm}
}
The \flatm control cannot express a conflict log. Its whole-string majority has no precaution field to attach dissent to, so the dissent is voted away. In the contradiction grid of Fig.~\ref{fig:contradiction}, we therefore report the combination of \flatm with a conflict log as not expressible.

\section{$\tau$-bench results}
\label{app:tau}

\begin{table}[htbp]
\centering\footnotesize
\begin{tabular}{@{}lccc@{}}
\toprule
Arm & step acc. & task succ. & steps \\
\midrule
\textsc{Plain} & .596\sd{.036} & .297\sd{.046} & 264 \\
\docs & .679\sd{.012} & .364\sd{.048} & 264 \\
\synapse & .663\sd{.025} & .370\sd{.010} & 264 \\
\central & .672\sd{.019} & .358\sd{.038} & 264 \\
\bottomrule
\end{tabular}
\caption{On $\tau$-bench retail the three compendium arms are within one SD of one another. 55 held-out tasks, GPT-4o (August 2024) agent and simulator, three seeds. \textsc{Plain} is our own agent loop in the vendored harness and is not comparable to the leaderboard.}
\label{tab:tau}
\end{table}

The primary metric is per-step tool-call accuracy (264 decisions per seed; Tab.~\ref{tab:tau}), because task success on 55 tasks cannot separate the arms. We score it against gold tool calls by longest-common-subsequence matching of tool names within each task. The three compendium arms lie within one SD of one another. We detect no added benefit from federated experience on 14 well-documented tools. We vendor the public $\tau$-bench harness and record its revision per run. The four arms ran under two revisions that differ only in a file none of the runs executes. A \docs replication on the later revision gave 0.664 / 0.339, within one SD of 0.679 / 0.364. \textsc{Plain} runs at pass\textasciicircum1 on a 55-task holdout.

\section{Why we do not use topic-labeled proxies}
\label{app:proxies}

Multi-agent routers are evaluated on math benchmarks such as GSM8k, where they route queries to agent roles \cite{yue2025masrouter,yang2025bmar}. A tool-routing proxy can be built on such benchmarks (GSM8k, BBH) by assigning each problem a ``tool'' label. We built such a proxy to test whether it measures routing. The labels are topics, not tools (Algebraic Word Problem Solver, Percentage and Proportion Solver, Financial and Banking Calculator, Work/Rate/Time Analyzer, Geometry and Measurement, General Logic and Counting, Number Theory). No two differ in what an executor can do. The gold label is thus a property of the annotator's taxonomy. Financial and Banking queries split exactly 56/56 between their own label and General Logic and Counting. Of 16 Geometry queries, 13 were routed to Work/Rate/Time. On 500 proxy queries, a TF-IDF (word 1--2-gram + char 3--5-gram) linear SVM reaches 0.92. It still reaches 0.82 when every scenario in the compendium is randomly relabeled. The router reaches 0.44 under four prompts (0.31--0.49), against a tool-level recall@5 of 0.89. We treat a routing benchmark as informative only under two conditions. The gold label must be determined by what the query needs done, and different labels must correspond to executors with different capabilities. StableToolBench, $\tau$-bench, and LiveMCPBench \cite{mo2025livemcpbench} satisfy both. Topic-to-family mappings satisfy neither.

\end{document}